\documentclass[twocolumn]{aastex631}

\usepackage{graphicx}
\usepackage{amsmath}	
\usepackage{subfigure}
\usepackage{wrapfig}
\usepackage{booktabs}
\usepackage{comment}
\usepackage{ulem}
\usepackage{xcolor}
\usepackage{multirow} 

\begin{document}


\shorttitle{RAMBO I}
\shortauthors{Z. Keszthelyi et al.}

\title{RAMBO I: Project introduction and first results with uGMRT}

\author{Z. Keszthelyi}
\correspondingauthor{Zsolt Keszthelyi}
\email{zsolt.keszthelyi@nao.ac.jp}
\affil{Center for Computational Astrophysics, Division of Science, National Astronomical Observatory of Japan, 2-21-1, Osawa, Mitaka, Tokyo 181-8588, Japan}

\author{K. Kurahara}
\affil{Mizusawa VLBI Observatory, National Astronomical Observatory Japan, 2-21-1 Osawa, Mitaka, Tokyo 181-8588, Japan}

\author{Y. Iwata}
\affil{Division of Science, National Astronomical Observatory of Japan, 2-21-1, Osawa, Mitaka, Tokyo 181-8588, Japan}
\affil{Mizusawa VLBI Observatory, National Astronomical Observatory of Japan, 2-12 Hoshigaoka, Mizusawa, Oshu, Iwate 023-0861, Japan}

%
%

\author{Y. Fujii}
\affil{Division of Science, National Astronomical Observatory of Japan, 2-21-1, Osawa, Mitaka, Tokyo 181-8588, Japan}
\affil{Department of Astronomical Science, School of Physical Sciences, Graduate University for Advanced Studies (SOKENDAI), 2-21-1 Osawa, Mitaka, Tokyo 181-8588, Japan}

\author{H. Sakemi}
\affil{Graduate School of Science and 
Technology for Innovation,
Yamaguchi University, 1677-1, Yoshida, Yamaguchi, Yamaguchi 753-0841, Japan}

\author{K. Takahashi}
\affil{Division of Science, National Astronomical Observatory of Japan, 2-21-1, Osawa, Mitaka, Tokyo 181-8588, Japan}
\affil{Department of Astronomical Science, School of Physical Sciences, Graduate University for Advanced Studies (SOKENDAI), 2-21-1 Osawa, Mitaka, Tokyo 181-8588, Japan}

\author{S. Yoshiura}
\affil{Mizusawa VLBI Observatory, National Astronomical Observatory Japan, 2-21-1 Osawa, Mitaka, Tokyo 181-8588, Japan}

\begin{abstract}

Magnetic hot stars can emit both coherent and incoherent non-thermal radio emission. Understanding the nature of these emissions and their connection to stellar rotation and magnetic field characteristics remains incomplete. 
The RAdio Magnetospheres of B and O stars (RAMBO) project aims to address this gap by systematically detecting and characterizing gyrosynchrotron and cyclotron maser radio emission in rapidly rotating magnetic hot stars. Using the upgraded Giant Metrewave Radio Telescope, we present the first detection of radio emission from HD55522 at 650 MHz, confirming it as a new radio-bright magnetic hot star. This supports the predictions of the Centrifugal Breakout model, furthering its application in understanding particle acceleration mechanisms in centrifugal magnetospheres of hot stars.
Additionally, we report non-detections for four other targets, improving sensitivity limits by a factor of a few compared to previous observations. These findings demonstrate the potential of RAMBO to uncover the complexities of radio emission in massive stars and highlight the need for broader, multi-wavelength observations to probe magnetospheric physics comprehensively. The sensitivity of the Square Kilometre Array will enable significant advancements. 

\end{abstract}


\section{Introduction} 
\label{sec:intro}

%
%
Massive stars play a vital role in feedback processes on local and even galactic scales and are essential in the chemical evolution of the Universe \citep[e.g.,][]{geen2023}. It has been conclusively demonstrated that some massive stars with OB spectral types possess large-scale, globally-organized magnetic fields \citep[e.g.,][]{fossati2015,wade2016,grunhut2017,shultz2018}. Since these fields do not show any apparent correlation with stellar parameters, notably stellar rotation, it is most likely that they are fossil fields without an active generating mechanism, whose exact origin still remains heavily debated \citep[e.g.,][for reviews]{donati2009,braithwaite2017,keszthelyi2023}. Such fossil fields have a significant impact on the evolution of stars by modifying their mass loss and aiding their spin-down \citep[e.g.,][]{petit2017,keszthelyi2019,keszthelyi2020,schneider2020,takahashi2021,takahashi2024}. 

%
%
Fossil fields form a magnetosphere around the star, typically extending to several stellar radii, which is determined by the ratio of magnetic energy density and stellar wind kinetic energy density. The most commonly inferred geometry is predominantly dipolar \citep[e.g.,][]{shultz2018}.
While hot star magnetospheres have been extensively studied both observationally and theoretically, critical uncertainties exist in describing plasma flow and redistribution in the magnetosphere, loss and leakage mechanisms, and characterizing the magnetic field geometry in greater detail. These have important implications for correctly interpreting observational characteristics and the magnetic field effects studied in stellar models.

%
%
Radio emission from hot stars can be categorized in terms of thermal (stellar wind-driven) and non-thermal (electron acceleration) emission. The latter has been found to display both a non-coherent (gyrosynchrotron) and coherent emission (auroral radio emission, ARE). 
ARE is powered by anisotropic electron velocity distributions in the stellar magnetosphere, producing electron cyclotron maser emission (ECME). 
These unstable distributions, such as loss-cone or ring distributions, provide the free energy necessary for wave amplification and can form through processes such as magnetic mirroring, where electrons are reflected by converging magnetic fields, and magnetic reconnection, which accelerates electrons and enhances anisotropies \citep{wu1979,melrose1982,winglee1986}. Additionally, the interaction of electron beams with ambient plasma can influence ECME properties, as beam dynamics affect the growth rates and modes of the emission. ECME is marked by high temperatures and significant circular polarization, making it an efficient tracer of magnetospheric dynamics.

%
%
The Centrifugal Breakout (CBO) model was first proposed to explain the H$\alpha$ emission from hot stars \citep[][]{owocki2020}. H$\alpha$ emission is uniquely observed from those stars that possess a centrifugal magnetosphere \citep[][]{petit2013}. In this case, stellar rotation supports the build-up of material above the Kepler co-rotation radius. Centrifugal breakout occurs when the plasma confined in the magnetosphere reaches a critical density. This opens up the magnetic field lines and releases the plasma.
Recognizing the correlation between H$\alpha$-bright and radio-emitting hot stars, the CBO model was extended to describe the origin of non-thermal, primarily gyrosynchrotron, radio emission \citep[][]{owocki2022}. In brief, the key new component of this model is to attribute the source energy of electron acceleration to stellar rotation. Indeed, observed samples show that radio emitters are in almost all cases rapid rotators, which challenges the concept of a purely wind-driven reconnection \citep[][]{leto2021,shultz2022}. While the CBO model provides a robust theoretical framework for understanding the origin of gyrosynchrotron emission, it requires scaling down the predicted CBO luminosity by a factor of $10^{-8}$ to match observed radio luminosities. This factor is currently justified as an empirical correction accounting for plasma heating, radiation losses, and other processes that dissipate most of the rotational energy before it contributes to non-thermal radio emission. However, the precise physical mechanisms governing this scaling remain poorly constrained and may vary between stars. Similarly, the flux-to-luminosity conversion of observed stars assumes a trapezoidal spectral energy distribution (SED) model \citep{shultz2022}, which integrates over a wide frequency range (0.6 GHz to 100 GHz) to approximate the total radio luminosity used in the CBO relation. While validated against multi-frequency observations, deviations in individual cases, such as stars with steep spectral turnovers, could introduce systematic uncertainties.

Several efforts have already been undertaken to study hot stars in the radio domain via modern instruments. 
Recently, \cite{leto2018} demonstrated that HR5907 (HD142184), a fast-rotating magnetic hot star, shows strong gyrosynchrotron emission from non-thermal electrons accelerated in the magnetosphere. From their radio observations, they also found evidence suggesting that the magnetic field of this star may be non-dipolar, indicating the potential complexity of magnetic field topologies in hot stars.
\cite{das2022a} reported wideband observations of HD35298, which allowed the identification of the upper cut-off frequency of its ECME spectrum and led to a review of existing scenarios, suggesting that multiple physical processes might explain the premature cut-off of ECME in hot magnetic stars. \cite{das2022b} conducted radio observations of five hot magnetic stars to test the robustness of the proposed scaling relation between ECME luminosity, magnetic field strength, and stellar temperature, discovering three new ECME-producing stars and concluding that while the relation remains valid, the temperature dependence for late-B and A-type stars may be less reliable. 
\cite{das2022c} significantly expanded the sample size of main-sequence radio pulse emitters by discovering eight additional objects, allowing for the first statistical analysis of their physical properties, and suggesting that at least 32 percent of magnetic hot stars exhibit ECME, with the primary factors being the maximum surface magnetic field strength and temperature.
\cite{hajduk2022} searched for non-thermal radio emission from chemically peculiar stars at 120-168 MHz using the Low-Frequency Array (LOFAR), and reported detections of two stars (BP Boo and $\alpha^2$ CVn), with findings supporting theoretical predictions of a turnover at low frequencies and reduced emission variability with rotational phase.
\cite{biswas2023} used simultaneous Band 4 (500-900 MHz) and Band 5 (1050-1450 MHz) observations by the upgraded Giant Metrewave Telescope (uGMRT) and L-band (900-1670 MHz) and UHF band (580-1015 MHz) observations by MeerKAT and discovered strong variable radio emission from the eccentric binary $\epsilon$~Lupi, the first high-mass, main-sequence binary system where both stars are magnetic. The emission characteristics suggest variable magnetic reconnection throughout the orbital cycle, making it a key system for studying magnetospheric interactions.
\cite{polisensky2023} searched for transient radio flares from 761 hot magnetic stars using the VLITE Commensal Sky Survey, detecting three potential sources and concluding that the data are consistent with the hypothesis that these flares originate from the stars' magnetospheres, although further data are needed for a definitive association.

%
%
However, despite these efforts, the characterization of hot magnetic stars in radio frequencies remains far from complete. On the one hand, several sources are expected to display gyrosynchrotron emission, but they remain undetected or unobserved. On the other hand, existing observations are often severely limited in frequency and time coverage, which can only be resolved with multiband observations and monitoring campaigns. To fill these gaps and to evaluate the newly proposed CBO model, improved data is urgently required. 
A critical limitation in the study of massive magnetic stars is the lack of phase-resolved radio observations over the rotation period and consistent long-term optical spectropolarimetric monitoring. Phase-resolved radio data allow the characterization of flux variability over the rotation cycle, which is crucial to understanding whether gyrosynchrotron or coherent emission dominates at specific phases. Similarly, optical spectropolarimetric monitoring enables precise determination of magnetic field geometries, including obliquity and alignment with the observer’s line of sight. Without these data, the interpretation of radio luminosities and non-detections remains inherently uncertain, especially for stars with complex magnetospheric dynamics.

In this work, we introduce a new project that aims at studying hot magnetic stars at radio frequencies, focusing on non-thermal emission. Our first results obtained with uGMRT are evaluated in the context of the CBO model. 
%
%
This paper is structured as follows. In Section~\ref{sec:cbo}, we outline the CBO model in more detail. In Section~\ref{sec:rambop}, we introduce the motivation and goals for the RAMBO project, our overall target selection, and the specific target selection for the uGMRT campaign. In Section \ref{sec:obs}, we describe the methods and observations. In Section \ref{sec:res}, we present our results from the first campaign and then in Section \ref{sec:disc} we discuss the implications and broader context of our results. Finally, in Section \ref{sec:concl}, we conclude and summarize our findings.

\section{The CBO model for radio emission} 
\label{sec:cbo}

\subsection{General considerations of the CBO model}
The Rigidly-Rotating Magnetosphere (RRM) model, introduced by \cite{townsend2005}, has been successfully applied to explain Balmer line and photometric variability in magnetic hot stars. A key component of the RRM model is stellar rotation, which supports the stellar wind material in the magnetosphere. The co-rotation of this material leads to the variability in observable diagnostics.
However, among several remaining questions, the mass removal from centrifugal magnetospheres is still an unresolved problem. Both leakage and breakout scenarios have been proposed \citep{owocki2018,owocki2020}. The Centrifugal Breakout model considers that when a critical plasma density is reached, the magnetic field lines cannot confine the plasma anymore and will open up to release this material. From analytical considerations, the characteristic surface density is cast in the form of:
\begin{equation}\label{eq:dens}
    \sigma_\star = \frac{B_K^2}{4 \pi G M_\star / R_K} \, , 
\end{equation}
with $B_K$ the magnetic field strength at the distance of the Kepler co-rotation radius $R_K$, $G$ the gravitational constant, and $M_\star$ the mass of the star. The MHD-modified scaling by \cite{owocki2020} invokes a stronger $R_K$ dependence.
The plasma build-up in the centrifugal magnetosphere is a function of the obliquity angle \citep{townsend2005}. The confined mass is maximum when the rotation and magnetic axes are aligned ($\beta=0^\circ$), whereas the confined mass is minimum when the two axes are perpendicular ($\beta=90^\circ$).
\cite{uddoula2023} performed detailed 3D simulations with oblique rotators, indeed finding plasma distribution in a warped disc or a ``wing" shape along the magnetic equator.

%
%
\subsection{Models for radio emission}

The gyrosynchrotron radio emission observed from the magnetospheres of early-type stars is attributed to relativistic electrons gyrating along magnetic field lines. Earlier models, such as those by \cite{trigilio2004} and \cite{leto2006}, proposed that these electrons are accelerated in the outer magnetosphere, where the stellar wind stretches magnetic field lines into a current sheet (CS) between regions of opposite polarity. In particular, \cite{leto2006} extended the CS framework by introducing a radiation belt-like structure, where wind-driven plasma accumulates and powers emission through electron spiraling. While these models provided a foundation for understanding magnetospheric radio emission, they assumed that the stellar wind dynamics alone drove electron acceleration, excluding any significant role for stellar rotation.

Subsequent observations, however, firmly established that the strength of radio emission correlates with stellar rotation rates\footnote{Though, empirically already evidenced in smaller samples \citep[e.g.,][]{linsky1992}.}, challenging wind-driven paradigms \citep{leto2021,shultz2022}. This realization led to the development of rotation-centric models, such as the CBO framework \citep{owocki2022}, which identifies stellar rotation as the primary driver of electron acceleration. In magnetic stars with centrifugal magnetospheres, plasma accumulates near the Kepler co-rotation radius, where centrifugal forces balance gravity. When the plasma density exceeds a critical threshold (Equation \ref{eq:dens}), magnetic tension can no longer confine the material, triggering centrifugal breakout events. These events lead to magnetic reconnection, releasing stored magnetic energy that accelerates electrons to relativistic speeds. These high-energy electrons then spiral along magnetic field lines, generating the observed gyrosynchrotron emission.

The CBO model offers several advantages over earlier approaches. It explains the observed strong dependence of radio luminosity on both the stellar rotation rate and magnetic field strength. Unlike the CS models, which rely solely on wind dynamics, the CBO framework places stellar rotation as the main energy source powering radio emission.
A key refinement to the CBO model emphasizes the quasi-steady-state nature of magnetospheres in these stars \citep{shultz2022}. Instead of sporadic, large-scale breakouts, the magnetosphere appears to self-regulate near the critical breakout density, allowing for continuous, small-scale centrifugal breakout events. This quasi-continuous process sustains the production of relativistic electrons through localized reconnection, maintaining steady gyrosynchrotron emission over time. The framework also reconciles the long-term stability of magnetic fields with their role as a conduit for rotational energy. Together, these models underscore the evolving understanding of magnetospheric processes, with CBO-driven reconnection emerging as a robust explanation for observed radio emission.

%
%
\subsection{CBO model predictions for radio luminosity}
The CBO model predicts that the observable non-thermal radio luminosity from a magnetic hot star should scale with the CBO luminosity $L_{\rm CBO}$, which can be cast in the form of: 
\begin{equation}\label{eq:cbo}
    L_{\rm CBO} = \dot{M} \Omega^2 R_\star^2 \eta_c^{1/p} \, , 
\end{equation}
with $\dot{M}$ the mass-loss rate, $\Omega$ the surface angular velocity, $R_\star$ the stellar radius, $\eta_c$ the centrifugal magnetic confinement parameter, and $p$ the multipole index (Eq. 13 of \cite{owocki2022}). For a dipole, $p=2$, and for a split monopole field, $p=1$. Interestingly, in the latter case, $\dot{M}$ cancels out and the CBO luminosity scales as $L_{\rm CBO} \propto B^2 R_\star^4 P_{\rm rot}^{-2}$, with $B$ the surface magnetic field strength and $P_{\rm rot}$ the rotation period of the star. This scaling was indeed found by \cite{leto2021} and supported by \cite{shultz2022}. Thus, it is empirically well-established that both magnetic field strength and rotational velocity scale positively with the observed radio luminosity\footnote{The rotation period, the inverse of the rotational velocity, scales with a negative power.}.
\subsection{CBO model successes and challenges}

The key success of the CBO model is the first, theoretical interpretation of stellar rotation as the energy source behind radio emission. 
However, Eq.~\ref{eq:cbo} is a purely theoretical quantity, which has been equated with the observable radio luminosity. Some corrections are therefore necessary to be incorporated. 
Firstly, \cite{shultz2022} and \cite{owocki2022} invoke a large $10^{-8}$ scaling down of the CBO luminosity to equate it with the radio luminosity. The origin of the scaling factor is only broadly attributed to plasma heating, which would consume most of the CBO luminosity. These contributions urgently need to be disentangled for more accurate predictions. 
Furthermore, some observationally-motivated corrections are yet to be described in the CBO model. On the one hand, \cite{shultz2022} and \cite{owocki2022} found that including the obliquity angle and re-introducing the mass-loss rate dependences improve the fit to the observed sample. 
The role of the obliquity angle has primarily been discussed in terms of plasma distribution within the magnetosphere, as described by RRM model. However, obliquity also influences the observed radio emission by affecting the alignment of the magnetosphere with respect to the observer. This effect is particularly evident in the rotational modulation of gyrosynchrotron radio emission, which varies over the entire rotation cycle. The observed modulation, typically of a factor of two to three, is relatively small compared to the order-of-magnitude increases associated with ECME. Notably, studies of stars with Oblique Rotator Model geometries—such as HD~37479 \citep{leto2012}, HD~182180 \citep{leto2017}, and HD~124224 \citep{das2021}—have demonstrated a clear correlation between the extrema and nulls of the magnetic curve and the maxima and minima of the total intensity radio emission (Stokes I), with an evident phase shift between the two light curves. These observations provide strong support for the impact of the obliquity angle on the observed radio emission.

While the mass-loss rates, somewhat surprisingly, vanish from the formal CBO scaling for a split-monopole, it is clear that self-absorption in the magnetosphere can attenuate radio emission. This correction, particularly to explain O-type stars, is yet to be quantitatively incorporated into the model.

Insofar, it remains to be explored whether magnetic hot stars would also emit non-thermal radio emission if they rotate slowly. This would point to a different mechanism powering the radio emission. 
Some works also discuss the potential to establish a more general scaling relation for ordered magnetospheres, including hot magnetic stars, ultracool dwarfs, planets such as Jupiter, and exoplanets \citep[e.g.,][]{leto2021,owocki2022}. 
However, before these broader goals can be investigated, it is crucial to confront the CBO model with improved and more extended data sets of magnetic hot stars and incorporate physically motivated roles of obliquity and mass loss. To further advance with these goals, we introduce the RAMBO project.


\section{RAMBO Project} 
\label{sec:rambop}

%
%
\subsection{Motivation and Goals} 
\label{sec:rambop1}

The detailed physical characterization of massive stars is essential to entangle the evolutionary pathways that lead to various kinds of supernovae, compact objects and their mergers, and hence astrophysical transients, such as gravitational waves, gamma-ray bursts and fast radio bursts (FRBs). The study of magnetic massive stars can provide a bridge to understanding the conditions that give rise to magnetars.
Recent discoveries suggest that magnetars are the origin of some FRBs, producing brief but intense bursts of radio waves \citep[e.g.,][]{andersen2020,bochenek2020,zanazzi2020}. 
Therefore, the characterization of their progenitors could improve the formation channel and population synthesis studies.
Radio observations of hot stars directly probe into their magnetospheres, where accelerated particles generate non-thermal emission. By extension, these observations may also offer complementary insights into the mechanisms behind magnetar-produced FRBs.

For these reasons, the ``RAdio Magnetospheres of B and O stars" (RAMBO) project aims to help currently ongoing efforts to study magnetospheric physics in hot ($T_{\rm eff} \sim 10 - 40$~kK), high-mass ($M_\star > 2$~M$_\odot$) stars. 
Primarily those radio facilities are used that have demonstrated capabilities of observing magnetic hot stars in the sub-GHz and few GHz bands, such as uGMRT, jVLA, and MeerKAT. 
Our explicit goal is to have a critical investigation of the origin and physical nature of radio emission from ordered magnetospheres, both on observational and theoretical grounds. The RAMBO project has three main objectives. 
\begin{enumerate}
    \item Detect gyrosynchrotron radio emission from those stars that are in the physically preferred parameter space according to the CBO model. 
    \item Monitor magnetic hot stars over their entire rotation periods and study the variability of radio flux. Identify ECME pulses and scrutinize their properties.
    \item Propose a theoretical explanation and critically evaluate the CBO model with improved data.
\end{enumerate}

%
%
%
\begin{figure*}
    \centering
    \includegraphics[width=0.97\textwidth]{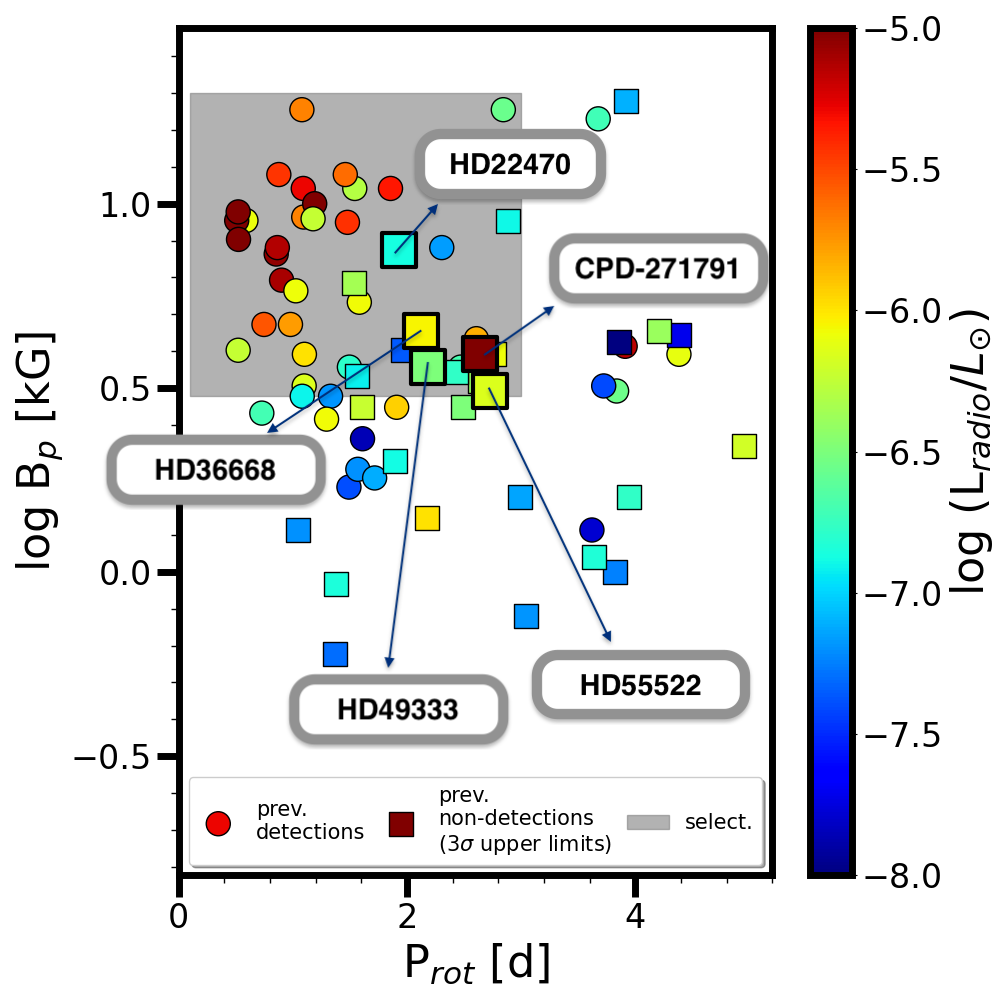} 
    \caption{Primary target selection for the RAMBO project on the polar magnetic field strength - rotation period plane. The data is based on the comprehensive collection of \cite{shultz2022}. The circles show previous radio detections, whereas the squares show non-detections. The colour-coding shows the radio luminosity (3 $\sigma$ upper limits in case of the non-detections). Our target selection for the uGMRT campaign of the physically best candidates according to the CBO model is shown with the grey area. The five stars reported in this paper are annotated.}
    \label{fig:targ}
\end{figure*}

%
%
\subsection{Overall Target Selection} 
\label{sec:rambo_targets}

\subsubsection{Primary Selection} 

Our overall target selection is aimed at confronting new observational data with the theoretical model of \cite{owocki2022} and is based on the previous comprehensive collection of \cite{shultz2022}. The \cite{shultz2022} catalog compiled all available radio measurements of magnetic early-type stars, resulting in the largest sample analyzed — 131 stars with rotational and magnetic constraints, including 50 radio-bright stars. The study confirmed a strong dependence of gyrosynchrotron radiation on stellar rotation and revealed a close correlation between H$\alpha$ emission strength and radio luminosity, suggesting that radio emission may be explained by the CBO mechanism. This catalog also challenges the previous paradigm of wind-driven electron acceleration, instead proposing that electrons are accelerated through centrifugally-driven magnetic reconnection, providing a unified explanation for both H$\alpha$ and radio emission from centrifugal magnetospheres.

Our sample is a directly adopted sub-sample from the \cite{shultz2022} catalog, albeit with updates when available. 
We cross-matched this catalog with the Sydney Radio Star Catalog \citep{driessen2024} and also reviewed lower-mass, Ap star samples, for example, \cite{hajduk2022}, to ensure that the catalog is as up-to-date as possible. 
To our knowledge, there is only 1 star, which is listed as a non-detection in the \cite{shultz2022} catalog; however, its radio emission has been since detected. This is HD36526, whose radio emission, in the context of ECME, was studied by \cite{das2022b}.

Figure~\ref{fig:targ} shows our overall target list, with previous detections and non-detections.
We focus only on those stars that have rotation periods of less than five days. This selection criteria is adopted since almost all hot magnetic stars with radio detection fulfil this condition and the CBO model specifically associates rotation with radio flux. From a practical point of view, the goal of (2) monitoring stars over their rotation period also becomes challenging when significant telescope time is required. Fast-rotating stars are clearly more favorable targets in this regard since the flux variability over the rotation period can be established with less telescope time.

In this parameter space, 44 stars have been identified as radio emitters and 29 stars have been observed but yielded no detection. According to the CBO model, all stars in this domain should show radio emission. However, several factors may have led to previous non-detections. We believe the primary reason is insufficient measurement sensitivity, which we aim to improve by at least a factor of a few and up to an order of magnitude. Therefore, as one of the main motivations of the RAMBO project, we aim to detect radio emission from best-candidate stars according to the CBO model. In Section \ref{sec:disc}, we discuss more comprehensively the potential technical and scientific reasons for non-detections.

\subsubsection{Secondary Selection} 

As secondary selection constraints, we consider three groups. 

\begin{itemize}
    \item \textbf{Magnetic stars without previous radio observations:} We performed a cross-match between the magnetic and radio observations, which yielded 16 stars in a similar parameter space as described above (a few stars have larger rotation periods than 5 days but still below 10 days) with available magnetic field measurements but no previous radio observations. These stars are thus straightforward candidates to confront the CBO model with new radio observations. 
    \item \textbf{Magnetic stars without H$\alpha$ emission or without centrifugal magnetospheres:} A few radio sources have been found outside of the parameter space outlined in the primary selection. While the vast majority of radio detections concern stars with centrifugal magnetospheres that show H$\alpha$ emission, radio emission was also found in stars that are not H$\alpha$ bright. For instance, HD 142990 (V913 Scorpii) is a magnetic B-type star that does not display significant H$\alpha$ emission yet has been detected as a radio source \citep{linsky1989,lenc2018,das2019}. Similarly, HD 133880 (HR Lup) lacks prominent H$\alpha$ emission but shows radio emission, indicating that radio detection is possible even in the absence of H$\alpha$ features \citep{lim1996,das2018,das2020}. Although the detections in both cases focused on coherent emission, associated with ECME, gyrosynchrotron radiation is also evidenced (e.g. Figure 3 of \citealt{das2020}). Stars without centrifugal magnetospheres showing radio emission would certainly point towards the limitations of the CBO model. This is yet to be investigated with new observations. 
    \item \textbf{Magnetic O-type stars:} Radio emission has been detected from some of the known magnetic O-type stars. However, these yielded thermal emission \citep[][]{kurapati2017}, which probes the wind-interstellar medium interaction at spatial scales larger than the magnetosphere. In the future, we aim to constrain if gyrosynchrotron emission could be detected from magnetic O-type stars. We will target the best candidates, which are the magnetic O-type stars with the lowest possible mass-loss rates.
\end{itemize}

%
%
\begin{table*}[t!]
\centering
\caption{Log of our uGMRT observations (ID: 45\_013, PI: Keszthelyi). All observations were conducted at a 650 MHz band with a 200 MHz total bandwidth (band 4), which is in line with previous detections using GMRT. The start of the observations is given in HJD, the heliocentric Julian date, and $t$ refers to the on-source time of the wide-band data.}\label{tab:log}
\begin{tabular}{llllll}
\toprule
\toprule
Star & Start Obs. & Date & $t$ [min.] &  Flux calibrator & Gain calibrator \\
\midrule
\midrule
CPD-271791  & 2460281.27083 & 3 Dec 2023  & 88 & 0706-231  & 3C138, 3C84   \\
HD 22470    & 2460282.22917 & 3 Dec 2023  & 40 &  0340-213, 0629-199 &   3C138, 3C84  \\
HD 36668    & 2460302.35417 & 24 Dec 2023 & 40 & 0522+012  & 3C138, 3C147, 3C84   \\
HD 49333    & 2460282.31250 & 4 Dec 2023  & 40 &  0340-213, 0629-199 & 3C138, 3C84  \\
\multirow{2}{*}{HD 55522}   & 2460281.39583 & 3 Dec 2023  & 40 & 0706-231  & 3C138, 3C147, 3C84   \\
            & 2460283.39583 & 5 Dec 2023  & 40 &  0706-231  & 3C138, 3C84    \\
\bottomrule
\end{tabular}
\end{table*}

%
%
\subsection{Target Selection for the uGMRT Campaign} 
\label{sec:rambop_ugmrt_targets}

For the uGMRT campaign, our goal is to detect radio emitters for the first time, that is, to identify gyrosynchrotron emission\footnote{It is also possible that in given phases, the stars would display ECME. This could only help the identification of these objects as radio emitters since ECME typically has an order of magnitude higher radio flux than gyrosynchrotron emission.} from magnetic hot stars. Our targets are in the "preferred" parameter space of the magnetic field strength-rotation diagram, therefore, on a theoretical basis, they should show radio emission. More specifically, from the overall target selection, we further constrain the parameter space to stars with magnetic field strengths above 3 kG and rotation periods less than 3 days (shown with grey area in Figure \ref{fig:targ}). According to the CBO model, the 12 stars with previous non-detections should be radio emitters. Their expected radio flux should be similar to other stars in this parameter range. 
We received time allocation to observe some of these targets with uGMRT. In the present paper, we discuss 5 stars from the first set of observations (ID: 45\_013, PI: Keszthelyi). Further observations and results will be presented in a forthcoming publication (Kurahara et al., in prep.).


%
%
\section{Observations} 
\label{sec:obs}

\subsection{uGMRT}

%

The uGMRT is located in Pune, India, and consists of 30 antennas. 
We conducted band~4 (550–750 MHz) observations of the targets. We specifically requested band~4 because this range has been proven successful in identifying radio bright sources \citep{chandra2015,shultz2022}. The observations were performed in parallel using both narrow-band and wide-band modes. We utilize the wind-band data.
The use of the GWB backend system of uGMRT allows for wider bandwidth data, in comparison to the older GSB backend system. This results in an improved signal-to-noise ratio (S/N) compared to earlier observations.
The central frequency was 650~MHz with a total bandwidth of 200~MHz. We obtained full-Stokes polarization data.  

The observations were carried out over five sessions from December 3 to 5 and December 24, 2023. Each scheduling block alternated between target objects. At the beginning and the end of each observing session, we observed 3C138 and 3C84 for 10 minutes as flux density, bandpass, and polarization calibrators. The gain calibrator, J0706-231, was observed for 5 minutes every 30 minutes. According to the observation logs, 26 or 27 antennas were used during the sessions. When the scheduled calibrators had already set, 3C147 was used as a substitute (applied only to the first sessions observing HD55522 and HD36668). The details of the observations are summarized in Table \ref{tab:log}.

\subsection{Data Reduction}

We used SPAM (Source Peeling and Atmospheric Modeling; \citealt{intema2017}), which is based on the Astronomical Image Processing System (AIPS), for data reduction. 
Although we obtained full Stokes polarization data, our analysis is focused on Stokes I, as currently SPAM only supports this component.
SPAM utilizes AIPS version 31DEC13 and is controlled via ParselTongue version 2.3 (with Python 2.7). Since there were several bright compact sources in the field of view, and the beam pattern of these bright sources was clearly visible in the dirty image, direction-dependent calibration (DDC; \citealt{intema2017}) was applied to improve the dynamic range of the final image. For direction-dependent calibration, the Tata Institute of Fundamental Research (TIFR) GMRT Sky Survey (TGSS; \citealt{intema2017}) source catalog was used. Imaging in SPAM was performed with a Briggs robustness parameter of -1.0. 


\subsection{Data Analysis}

For the analysis of the wide-band data, the dataset was divided into 4 sub-bands from 550 MHz to 750 MHz with 50 MHz increments. Each sub-band was analyzed individually and then combined using WSClean (w-stacking clean; \citealt{offringa2014}) to produce the full-bandwidth (200 MHz) radio image, centered at 650 MHz. Since all the targets were point sources located at the center of the field of view, advanced techniques such as multi-scale cleaning and primary beam correction were not employed. We used auto-mask and no negative options. For HD22470, lower bandwidth data was used because RFI severely impacted the observations. This means that the effective integration bandwidth was halved compared to the other targets.
Since the stars are effectively point sources, i.e., the beam size is comparable to the measurement area, we equate the intensity measurement ($I$ [Jy/beam]) with the flux density at the central wavelength ($F_{\nu=650 \rm MHz}$ [Jy]).

To localize our targets in the sky, we used publicly available optical and infrared data from SDSS and 2MASS, respectively. The fits files were downloaded from the SkyView portal\footnote{\url{https://skyview.gsfc.nasa.gov/}}. Since the targets are bright stars, they are saturated in the SDSS images. Therefore we use the 2MASS K-band images to determine the area within which we measure the radio flux. Upon inspecting the images, we also used the SIMBAD Astronomical Database\footnote{\url{https://simbad.cds.unistra.fr/simbad/}} to cross-match the radio observations with any known radio sources in the field of view.

The H$\alpha$ profiles provide information on the centrifugal magnetospheres, while the Stokes $V$ parameter reflects on the magnetic characteristics.  
Therefore, with the goal of visual inspection of the H$\alpha$ and Stokes $V$ profiles, we also reviewed optical spectropolarimetric observations of our targets, which were retrieved from the PolarBase portal \footnote{\url{https://polarbase.irap.omp.eu/}} \citep{petit2014}.

%
%

%
%
\begin{table*}[t!]
\centering
\caption{Observed stars, GAIA DR2 distances, calculated CBO luminosity, measured radio flux density (with 3$\sigma$ upper limits for non-detections), inferred radio luminosity from the CBO model, inferred radio luminosity from the measured flux.}\label{tab:res}
\begin{tabular}{lccccc}
\toprule
\toprule
Star &  $d$ [pc]  & $\log (L_{\rm CBO} / L_\odot) $ &   $F_{\rm obs, \nu=650 MHz}$ [$\mu$Jy] & $\log (L^{\rm CBO}_{\rm rad} / L_\odot)$ & $\log (L^{\rm obs}_{\rm rad} / L_\odot)$ \\
\midrule
\midrule
CPD-271791                & 1280  & 2.50  & $<$ 91  & -5.50 & $<$ -5.52\\
HD 22470                  & 118 & 2.82    & $<$ 657  & -5.18 & $<$ -6.74\\
HD 36668                  & 393 & 2.31    & $<$ 92   & -5.67 & $<$ -6.54\\
HD 49333                  & 217  & 2.17   & $<$ 159  & -5.83 & $<$ -6.82\\
\multirow{2}{*}{HD 55522} & \multirow{2}{*}{278} & \multirow{2}{*}{2.03} &  164 $\pm$ 26    & \multirow{2}{*}{-5.97} & -6.59\\
                          &                      &                       &  $<$ 96 &                      & $<$ -6.83 \\
\bottomrule
\end{tabular}
\end{table*}


%
%
%
%
\begin{figure*}
    \centering
    \includegraphics[width=0.33\textwidth]{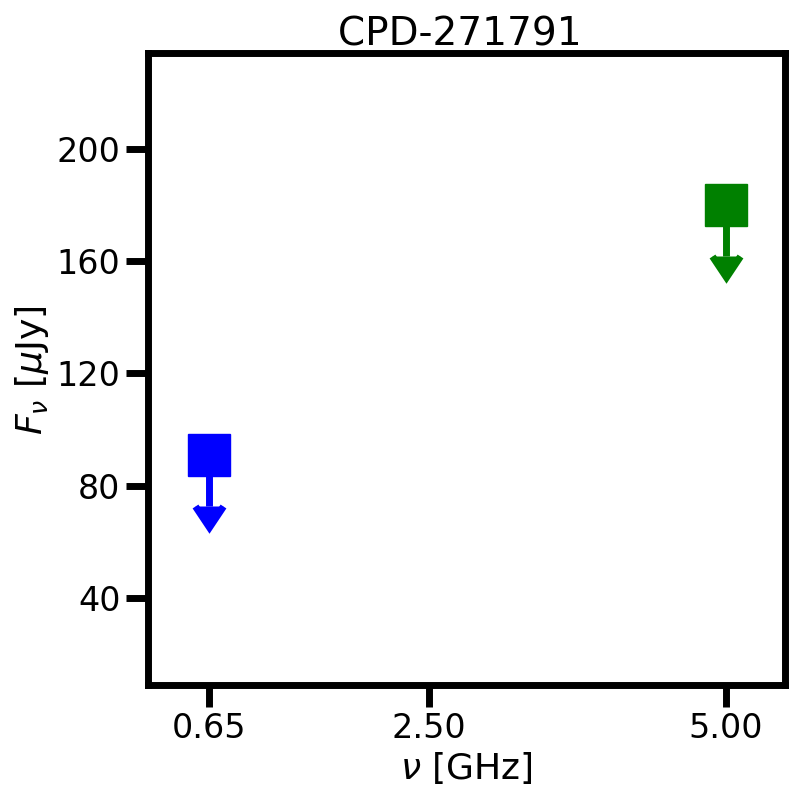}\includegraphics[width=0.33\textwidth]{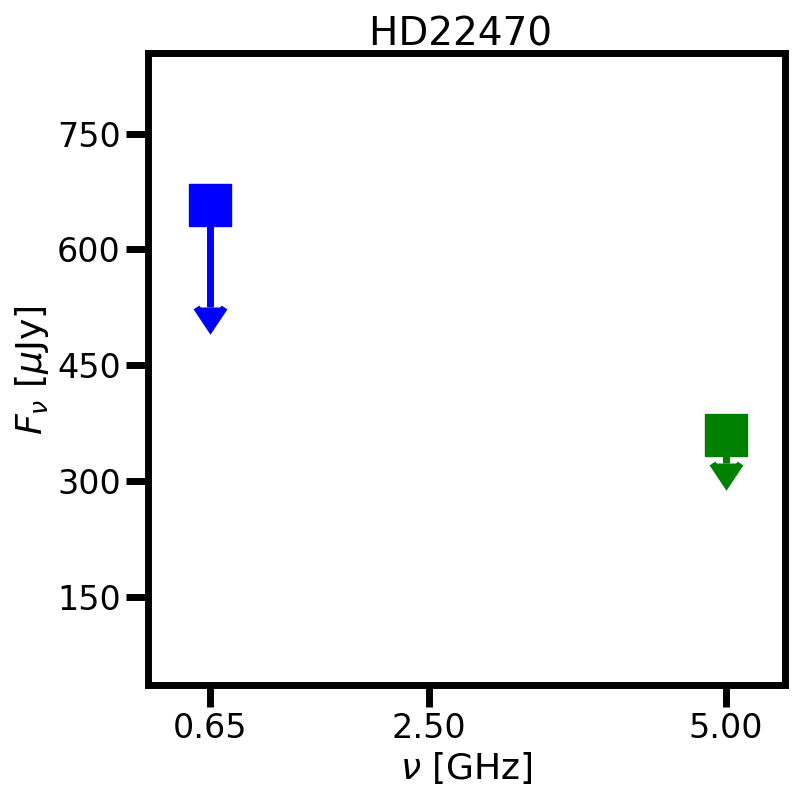}\includegraphics[width=0.33\textwidth]{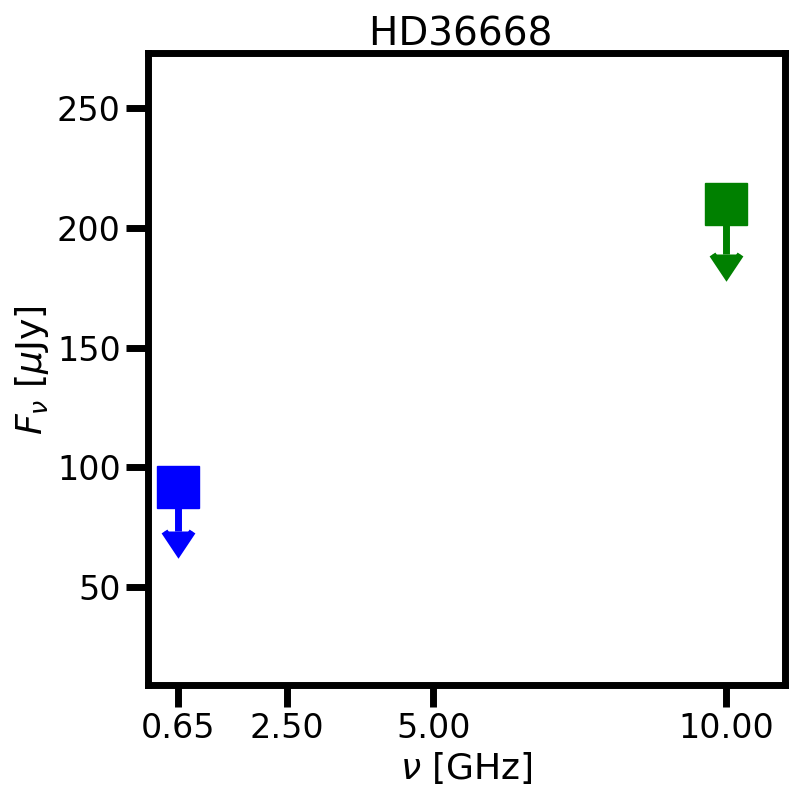}   
    \includegraphics[width=0.33\textwidth]{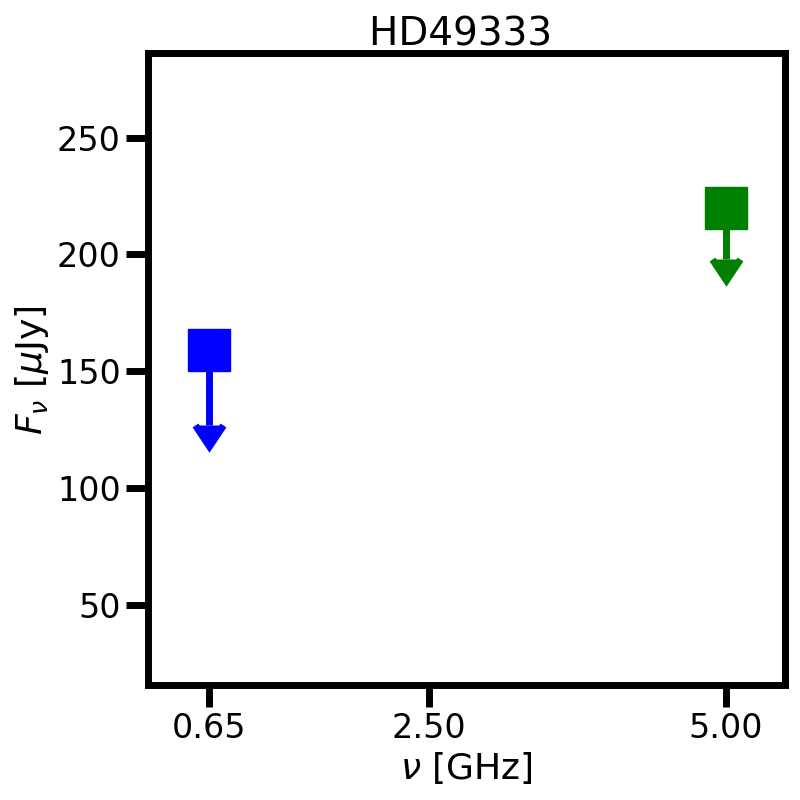}\includegraphics[width=0.33\textwidth]{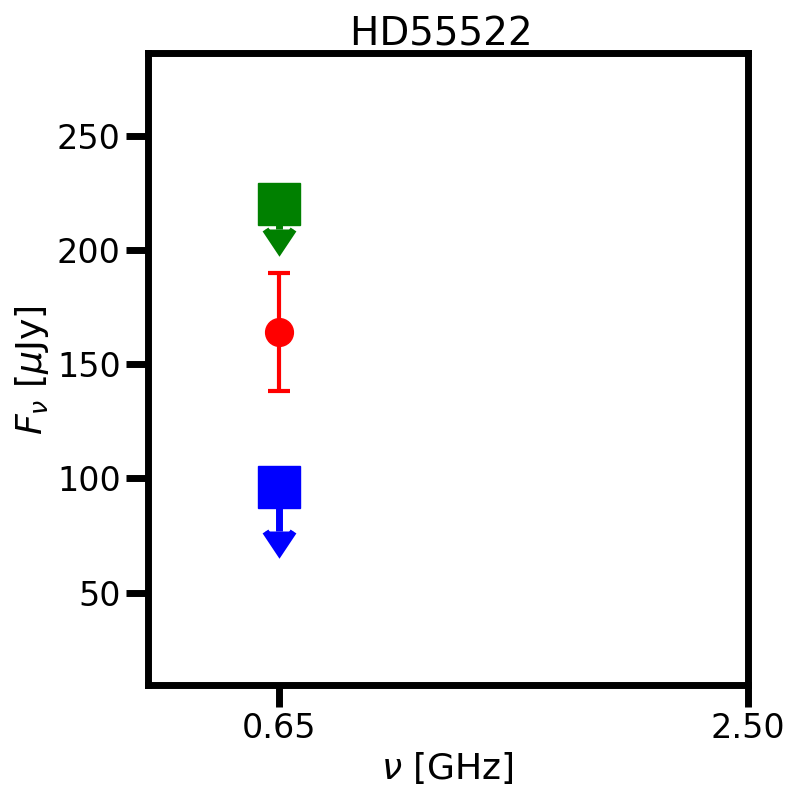}  
    \caption{Flux density in microJansky versus the frequency for our new observations (blue squares for non-detections, red circle for detection) and earlier measurements (green squares, all non-detections). The non-detections show the 3$\sigma$ upper limits. The size of the down-ward arrow is arbitrarily set. }
    \label{fig:fnu_obs}
\end{figure*}

\section{Results} 
\label{sec:res}

In this section, we present the analysis of five targets observed during our first uGMRT campaign. For each star, we summarize relevant literature data, including key stellar parameters and prior magnetic field measurements. We then discuss previous radio observations and provide results from our current radio observations. 
The images are shown in Figure~\ref{fig:rad1} and Appendix Figures~\ref{fig:star1_loc_rad}-\ref{fig:star5_loc_rad} and the measurement results for each star are shown in Figure \ref{fig:fnu_obs} along with the previous observations. Table~\ref{tab:res} summarizes the luminosity estimates. Here, we take the assumption that the expected radio luminosity from the CBO model follows the $10^{-8}$ scaling down and that the SED follows a trapezoidal function to calculate the radio luminosity from the flux density measurements.

%
%
%
\subsection{CPD-271791} 
\label{sec:star1}
%
%
CPD-271791 is a helium-strong early B-type star \citep[][]{garrison1977,walborn1983}. The effective temperature determined by \cite{zboril1999} and \cite{shultz2022} are $T_{\rm eff} = 22.7$~kK and $T_{\rm eff} = 23.8\pm1.6$~kK, respectively.
The magnetic field measurements were reported by \cite{jarvinen2018}. Since the optical spectropolarimetric data does not cover the entire rotation period of the star, the line-of-sight magnetic field strength indicates a minimum value of the assumed dipolar magnetic field with $B_{\rm p} > 3.9$~kG. We adopt this value in our calculations, keeping in mind that the real value might be higher and that the geometry might differ from a pure dipole. The rotation period was determined by phasing the optical photometric variability. This yielded $P_{\rm rot} = 2.641$~d \citep{jarvinen2018}.  
The star is reported as an H$\alpha$-bright source by \cite{shultz2020}. They calculated a Kepler co-rotation radius of $R_{K} = 4.1\pm0.2$~R$_\star$. Although, the obliquity angle (tilt of the magnetic axis compared to the rotation axis in a clockwise direction), $\beta $, cannot be constrained from the single spectropolarimetric data, \cite{shultz2020} argue that based on the shape of the H$\alpha$ profile, a high obliquity angle is likely. Therefore we assume $\beta=90^{\circ}$ in our calculations. Furthermore, based on previous inferences on the projected rotational velocity and rotation period, \cite{shultz2020} infer an inclination angle of the rotation axis compared to the observer of $i = 35\pm10^{\circ}$.

Previously, radio observations with VLA were conducted in the 5 GHz range \citep{drake1987,linsky1992}. The non-detection is constrained with a 180~$\mu$Jy upper limit on the radio flux. To our knowledge, there has not been newer radio observation specifically focused on this star. In particular, the GMRT archive search revealed no previous observation of this target.

%
%


%
%

From our observations with uGMRT at a central frequency of 650~MHz, we conclude a non-detection since the root-mean-square (rms) flux density is below the standard deviation. To provide a 3$\sigma$ upper limit, we report the value of the standard deviation times three. This standard deviation is measured in a radio-quiet area near the source (within 2 arcmin of the source in the 60 arcmin-size image). 
The upper limit is 91 $\mu$Jy.

For CPD-271791, the absence of long-term optical spectropolarimetric monitoring means that the full rotation phase curve is unconstrained. As a result, the inclination and obliquity angles cannot be reliably inferred, limiting the ability to model the magnetospheric geometry. Similarly, phase-resolved radio observations are unavailable for this star, precluding an assessment of variability over the rotation cycle, which may include critical features such as variability in gyrosynchrotron emission or peaks associated with auroral radio emission.

\subsection{HD22470} 
\label{sec:star2}
%
%
HD22470 (20 Eri) is a chemically peculiar (Si) Bp star. \cite{netopil2008} obtained an effective temperature of $T_{\rm eff} = 13.8\pm0.3$~kK, while \cite{paunzen2021} found $T_{\rm eff} = 12.6\pm0.3$~kK. 
The magnetic field of the star was first measured by \cite{borra1983}. Given the coverage of the optical spectropolarimetric data, a line-of-sight magnetic field phase curve could be established \citep{bychkov2021}. \cite{shultz2020} found a dipolar magnetic field strength of $B_{\rm p} = 7.5$~kG. Based on photometric observations, the rotation period of the star was constrained and refined by several authors over the years \citep{renson1981,adelman1995,adelman2000,dubath2011,bernhard2020,pyper2021}. \cite{shultz2020} found that the $P_{\rm rot} = 1.929$~d rotation period provides a good phasing for both the photometric and spectropolarimetric data. 
\cite{shultz2020} calculate an inclination angle of $i = 44^{\circ}$. A high obliquity angle is inferred for this star with $\beta=87^{\circ}$ \citep{glagolevskij2010,shultz2020}.

5-GHz radio observations of HD22470 are reported by \cite{drake1987} and \cite{linsky1992}. The non-detection is constrained with an upper limit of 360 $\mu$Jy. The GMRT archive search yields one previous observation that contained this star in the field of view with 18 min on-source time at 325 MHz frequency band (ID: 14HSA01, PI: Shukla). 

%
%
At the position of the sky associated with the target according to the 2MASS image, we find that the rms radio flux is below the standard deviation. Therefore, from our uGMRT observations, we report a 3$\sigma$ upper limit of 657 $\mu$Jy for a non-detection in HD22470.

\subsection{HD36668} 
\label{sec:star3}
%
%
HD36668 (V* V1107 Ori) is a helium-weak chemically peculiar (Si) B-type star \citep{topilskaya1993,adelman2000b} with an effective temperature of $T_{\rm eff}=$13.5~kK \citep{romanyuk2017}. 
The first magnetic field measurements were obtained by \cite{borra1981}. The line-of-sight magnetic field strength shows a quasi-sinusoidal variation. The dipolar field strength is estimated to be 4.5 kG \citep[][]{romanyuk2021,shultz2022}. Notably, \cite{shultz2022} infer a magnetic field strength that may contain a significant quadrupolar contribution. 
From TESS light curve analysis, \cite{romanyuk2021} obtained  $P_{\rm rot} = 2.1204$~days. Since this rotation period does not provide a coherent phasing for the line-of-sight magnetic field variability,
\cite{shultz2022} also inspected the TESS light curve and 
refined the rotation period to $P_{\rm rot} = 2.1192$~days, concluding that further observations are required to better constrain both the magnetic field strength and the rotation period of this star.
\cite{shultz2022} reported an obliquity angle of $\beta=80^{\circ}$. 
From the known stellar and rotational parameters, we estimate an inclination angle of $i = 60^{\circ}$. 

According to \cite{shultz2022}, Drake et al. constrained the radio flux with an upper limit of 210 $\mu$Jy at 10 GHz. 
While GMRT observations that contain HD36668 in the field of view exist (ID: 27\_048\, PI: Chandra and ID: 40\_069, PI: Bhattacharyya), to our knowledge, they were not analyzed and reported previously in the context of aiming to detect radio emission from this target. 
%
%
We report a non-detection with a 3~$\sigma$ upper limit of 92~$\mu$Jy for HD36668 from our new uGMRT observations.

%
%
\begin{figure*}
    \centering
    \includegraphics[width=0.45\textwidth]{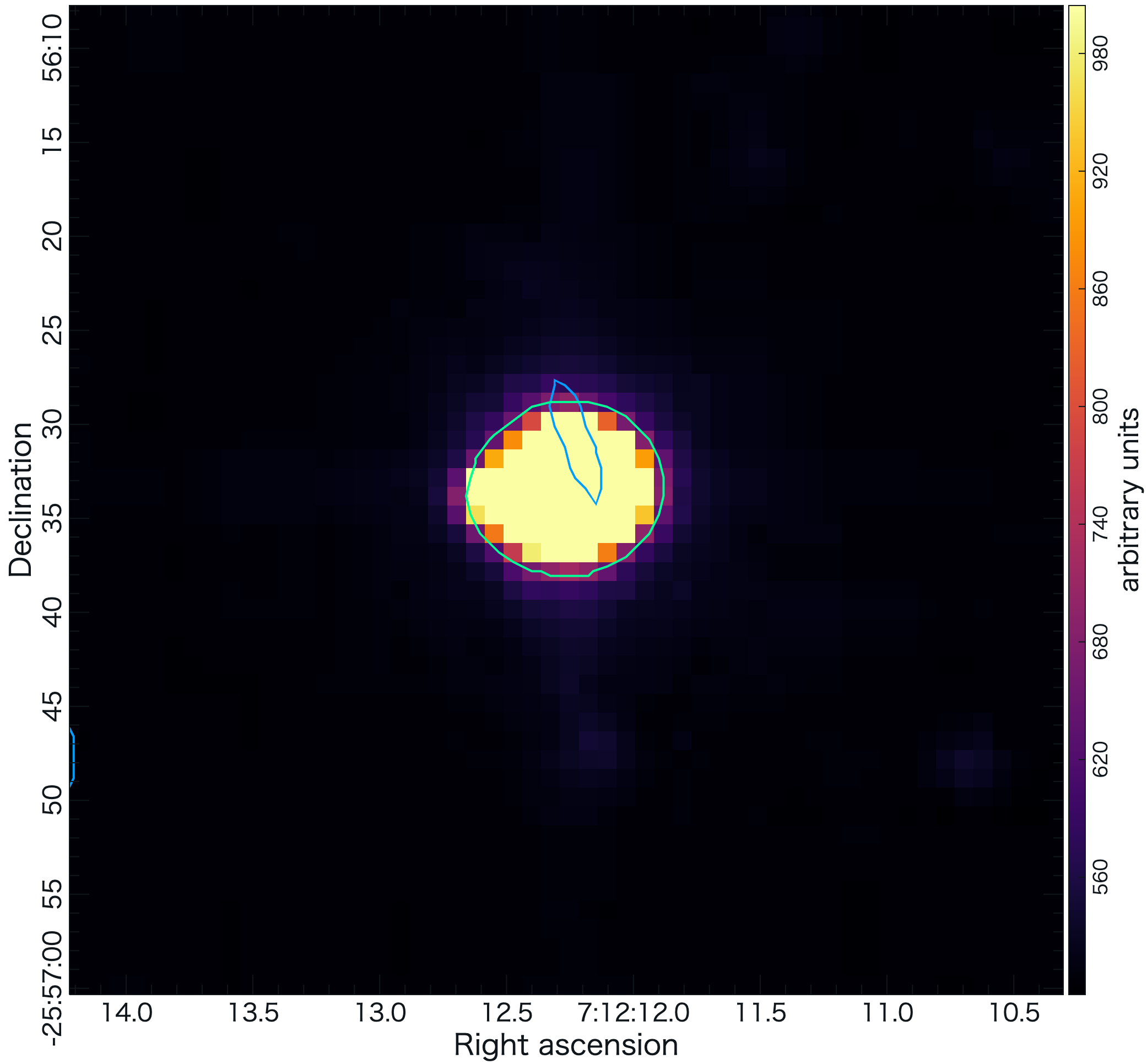}\includegraphics[width=0.45\textwidth]{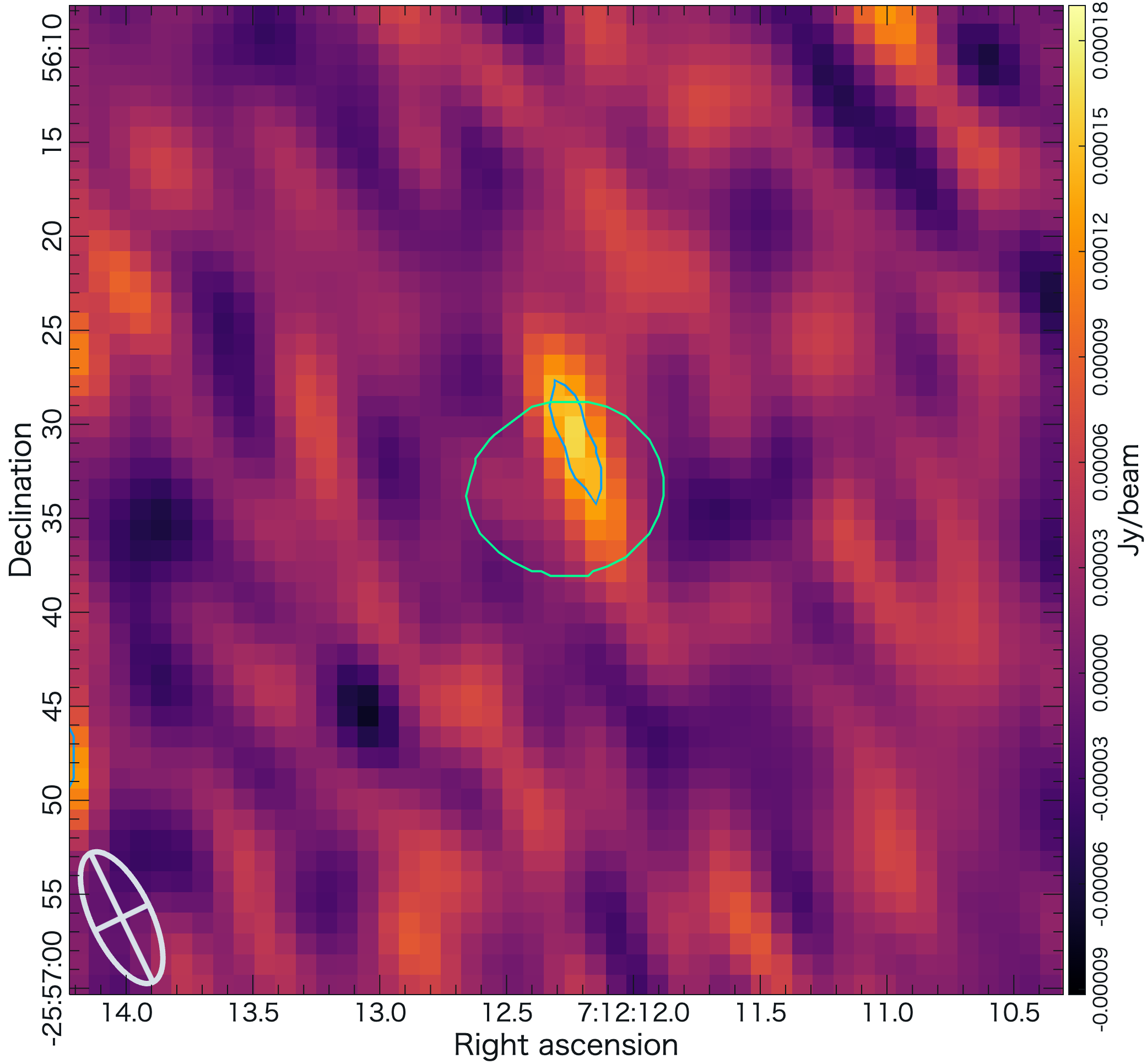}   
    \caption{K-band infrared image from 2MASS (left) and our uGMRT radio observation from Dec 3, 2023 of HD55522 (right). The beam size is shown in the lower left of the radio image. The blue contours of the radio intensity are at a $5\sigma$ level. Radio emission originating from the star's position in the sky is clearly evidenced. }
    \label{fig:rad1}
\end{figure*}

\subsection{HD49333} 
\label{sec:star4}
%
%
HD49333 (12 CMa) is an $\alpha^2$ CVn variable, helium-weak Bp star with an effective temperature of $T_{\rm eff}=15.8\pm0.1$kK \citep[][]{netopil2008}. It shows variable helium and silicon lines \citep[e.g.,][]{pedersen1977,farthmann1994,bailey2013,monier2023}. 
The star's magnetic field has been studied by \cite{borra1983,bohlender1993}. \cite{shultz2022} concluded a dipolar magnetic field strength of $B_{\rm p}= 3.6$~kG. 
A rotation period of $P_{\rm rot}=2.180$~d was constrained by several authors \citep{pedersen1979, bohlender1993,adelman2000,renson2001}. \cite{shultz2022} used this rotation period to phase the line-of-sight magnetic field variability \citep[see also][]{bychkov2021}, and found a magnetic obliquity angle of $\beta=85^{\circ}$. This is consistent with the discussion of \cite{bohlender1993}. While \cite{farthmann1994} estimated an inclination angle between 25 and 55$^{\circ}$, based on the newer stellar and rotational parameters we obtain $i = 65^{\circ}$. 

\cite{linsky1992} conducted radio observations of the star, finding an upper limit of 220 $\mu$Jy for the non-detection at 5 GHz. There is no record of observations of this target in the GMRT archive. 
%
%
From our uGMRT observations, we find a non-detection with a 3$\sigma$ upper limit of 159 $\mu$Jy.

\subsection{HD55522} 
\label{sec:star5}
%
%
HD55522 (*26 CMa) is a helium-strong B-type star with an effective temperature of $T_{\rm eff}=17.4\pm0.4$kK \citep[][]{briquet2004,petit2013}. The radial velocity variations hint at pulsations rather than binarity \citep{decat2002,gontcharov2006,aerts2014,bodensteiner2018}). Recently, Stochastic Low-Frequency variability \citep{bowman2019,bowman2022} was also reported in this star \citep{shen2023}.
The line-of-sight magnetic field variations are well-established \citep{kochukhov2006,bagnulo2015}, and indicate a dipolar structure with strength of $B_{\rm p}=$3.1~kG \citep{shultz2018,shultz2019a}.
A rotation period of $P_{\rm rot}=2.7292$~d is obtained, which fits well the optical photometric and spectropolarimetric datasets \citep[][]{briquet2004,shultz2018}. The inclination angle reported by \cite{briquet2004} is $i=80\pm10^{\circ}$, while \cite{shultz2019a} find $i=61\pm8^{\circ}$. We adopt the latter value. A magnetic obliquity angle of $\beta=89^{\circ}$ is found by \cite{shultz2019b,shultz2022}. 

X-ray observations yielded no detection in this star \citep{naze2014}.
Previous radio observations were obtained with GMRT (ID: 25\_045, PI: Bhattacharyya and IDs: 27\_048, 28\_075, PI: Chandra).  
\cite{shultz2022} analyzed the observations from 2015 September 15 (HJD: 2457280.57) and concluded a non-detection with an upper limit of 223 $\mu$Jy (in their Table B1, but 370 $\mu$Jy in their Table D1).

%
%
We observed this target two times in December 2023. 
%
On December 3, 2023, we observed HD55522, which led to detection, with the rms flux density higher than three times the standard deviation in the measured region. The region was determined from the K-band 2MASS image. We detect radio emission of $F_{\nu=650 \rm MHz} = (164 \pm 26)$ $\mu$Jy, and we associate it with the star (Figure~\ref{fig:rad1}). We determine the flux density as the peak within the measured region (see Figure \ref{fig:rad1}), while the uncertainty is adopted as the standard deviation of the image. Given the observation and analysis strategy, the uncertainty for radio detections is expected to be of the order of 10\% of the measured flux \citep{kurahara2023}. Indeed, adopting the worst-case consideration, that is, considering the standard deviation of the image as the measurement uncertainty, the uncertainty is 16\% of the measured flux density.

On December 5, 2023, we re-observed this star. This led to a non-detection, which we constrain with a 3$\sigma$ upper limit of 96 $\mu$Jy. This is consistent with the expectation that the gyrosynchrotron emission weakens by a factor of a few between different rotational phases. We further elaborate on this in Section~\ref{sec:disc} regarding data quality and phase constraints.


%
%
\section{Discussion} 
\label{sec:disc}

The results of our uGMRT campaign provide new insights into the nature of radio emission in magnetic hot stars and its relationship to the CBO model. We confirm the first detection of gyrosynchrotron radio emission from HD55522 at 650 MHz. Non-detection of the same target at a second epoch, as well as, non-detections for four other targets in single epoch observations highlight potential challenges in reconciling observations with theoretical expectations.
In this section, we first evaluate these findings in the context of the CBO model, considering both the implications of the detection and the possible reasons for non-detections, including observational sensitivity and model assumptions. We also examine the CBO-radio luminosity relation and explore how these results refine the next steps to help understand stellar magnetospheres and their emission mechanisms.
Then we discuss the measurement sensitivity, future prospects with SKA, and broader implications.

\subsection{Evaluation in the context of the CBO Model}

\subsubsection{Shultz et al. SED assumptions}

To estimate the radio luminosity from the measured flux densities of our targets, we first used GAIA parallaxes to obtain the distances. Then, we followed the conversion approach used by \cite{shultz2022}, which is based on and motivated by findings of \cite{leto2021}, described below. 

\cite{leto2021} showed that the radio spectra of magnetic hot stars can vary across a wide frequency range (from 0.1 to 400 GHz), depending on the properties of the star’s magnetosphere. By modeling the radio emission at different frequencies, they demonstrated that the shape of the spectral energy distribution (SED) can provide insights into the strength and structure of the magnetic field, as well as the distribution of high-energy electrons. Typically, the SED shows a flat profile over a wide frequency range, simplifying the conversion of measured radio fluxes to luminosity. This approach allows for a more accurate prediction of radio luminosities, even when only a limited number of flux measurements are available, by accounting for how radio emission behaves at different frequencies. This approach was used to estimate radio luminosities from single-frequency observations. To account for the broad range of possible frequencies, \cite{shultz2022} applied a trapezoidal function to integrate between 0.6 GHz and 100 GHz. The integration assumes flat emission between 1.5 GHz and 30 GHz, with the values set to unity, and drops off to zero at the extrema. 
Given these considerations, we obtain the frequency-independent luminosity as:
\begin{equation}\label{eq:1}
    L_{\rm radio} = 4 \pi d^2 \, F_{\rm peak} \Delta \nu \, , 
\end{equation}
with $d$ the distance, $F_{\rm peak}$ the peak flux density in the flat part of the SED, and
\begin{equation}\label{eq:2}
   \Delta \nu = \frac{1}{2} \left( (100 - 0.6) + (30 - 1.5) \right) \, [\mathrm{GHz}] \, .
\end{equation}

The explicit dependence on the SED can be omitted in the frequency-dependent, spectral radio luminosity $L_{\rm radio, \nu}$ \citep{leto2021,leto2022}. However, as long as the observations are in the flat part of the SED or there is no multi-wavelength data to confirm the peak flux $F_{\rm peak}$, the conversion between $L_{\rm radio}$ and $L_{\rm radio, \nu}$ is only a scaling by a constant factor of $\Delta \nu$. This is of the order of $10^{11}$~Hz \citep{leto2022}. Indeed, $\Delta \nu = 6.4 \cdot 10^{10}$~Hz, when using the \citealt{shultz2022} SED as specified above in Equation~\ref{eq:2}.

When multi-frequency observations are unavailable, \citet{shultz2022} assumed that the flux density measured at a single frequency corresponds to the peak of the SED, regardless of the frequency. While this approximation simplifies the calculation of radio luminosities, it leads to a factor of few uncertainties in the CBO luminosity prediction. 

Formally, this SED approach implicitly assumes zero flux density at the low-frequency end (0.6 GHz), which is inconsistent with observations showing emission at these frequencies (e.g., Table D1 of \citealt{shultz2022}). Evidence suggests that the flux density at 0.6 GHz is closer to the flat profile observed in the GHz range, indicating a need for refinement in the SED model.

We adopt the same assumption that our single-frequency measurements at 650 MHz --in lack of detection in the GHz range-- represent the peak of the SED ($F_{\rm \nu=650MHz} = F_{\rm peak}$). This aligns our analysis with \citet{shultz2022}, ensuring consistency in the derived radio luminosities and enabling direct comparisons with their results. While this method provides a practical solution for stars lacking multi-frequency data, it highlights the importance of future observations across broader frequency ranges to constrain SED models. Expanding sample sizes with multi-band data will enable refinements, better constraints on converting flux densities at sub-GHz frequencies to luminosity, and thus improve the predictive power of the CBO-radio luminosity scaling relation.

%
%
%
%
\begin{figure*}
    \centering
    \includegraphics[width=0.95\textwidth]{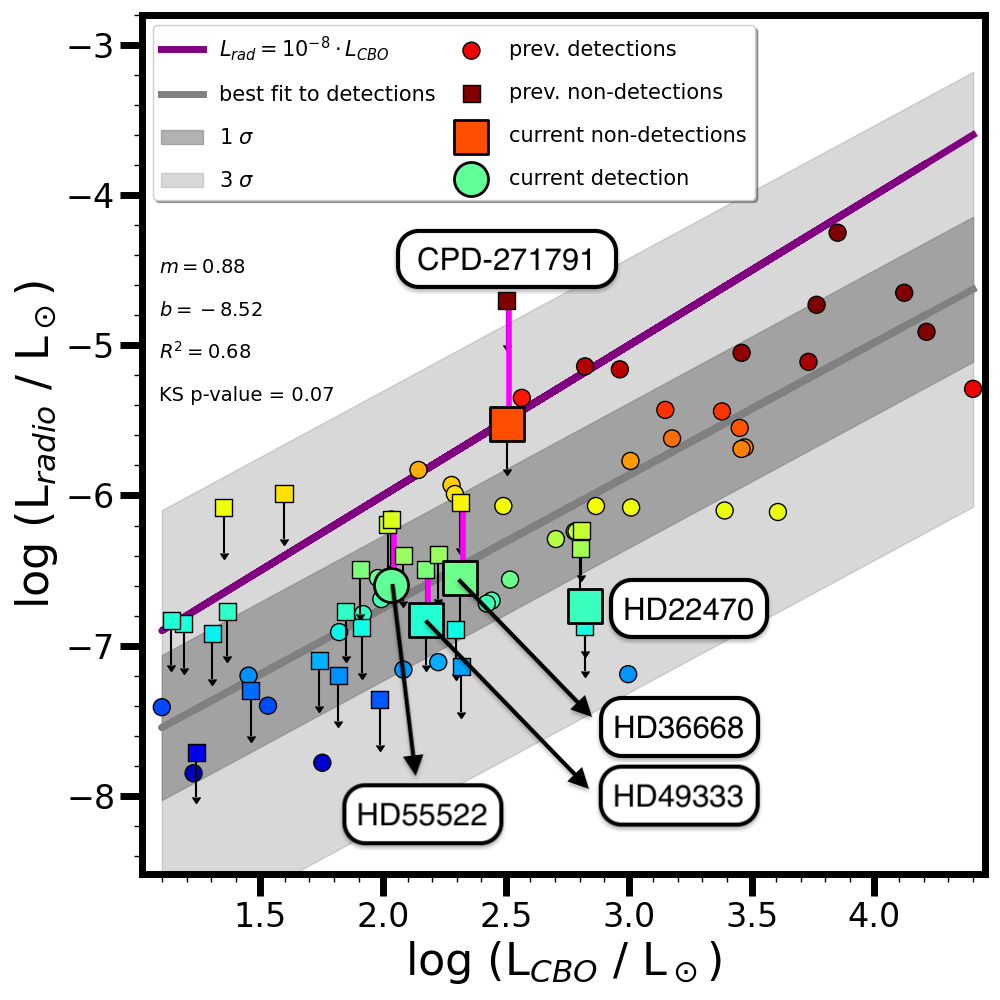}
    \caption{Radio luminosity versus the CBO luminosity, including a theoretical scaling (purple line), the best fit linear regression to radio-emitting sources (grey line), the 1 and 3$\sigma$ uncertainties of the regression (darker and lighter grey shaded areas). The bolometric solar luminosity is used as a constant scaling term for both axes. Previous detections are shown with smaller circles. Previous non-detections (indicating 3$\sigma$ upper limits) are shown with small squares. Our current observations are shown with the larger symbols, with connecting magenta lines to the previous observations for each target. The current non-detections are shown with larger squares and the current detection is shown with a larger circle. The color-coding is the same as in Figure~\ref{fig:targ}, showing the radio luminosity. The current non-detection for HD22470 almost overlaps with the previous non-detection.}
    \label{fig:lcbolrad}
\end{figure*}
%
%
%

%
%
%
%
\begin{figure}
    \centering
    \includegraphics[width=0.45\textwidth]{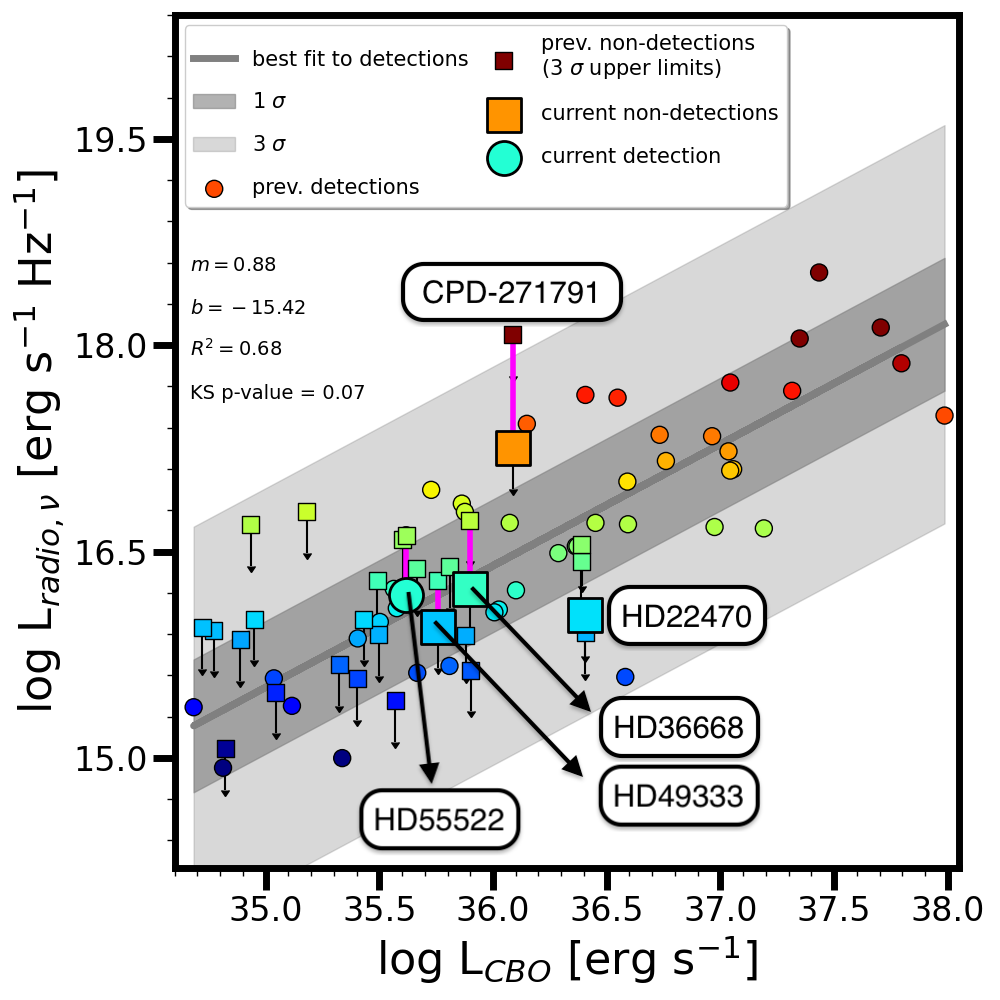}
    \caption{Frequency independent spectral radio luminosity vs the CBO luminosity in cgs units. The trends are identical to the ones on Figure~\ref{fig:lcbolrad}; however, the offset is different.}
    \label{fig:lcbolrad2}
\end{figure}
%
%
%

%
%
\subsubsection{CBO and radio luminosity correlation}

Figure~\ref{fig:lcbolrad} shows the correlation between CBO luminosity (\( L_{\rm CBO} \)) and radio luminosity (\( L_{\rm radio} \)). For this demonstration, to facilitate a direct comparison with \cite{shultz2022} and \cite{owocki2022}, we keep the units the same as in those works, and assume the split-monopole case with $p=1$. In this case, Equation~\ref{eq:cbo} simplifies to the form of:
\begin{equation}
  L_{\rm CBO} = \Omega^2 R_\star^4 B_{p}^2 v_{\rm orb}^{-1}  \, ,
\end{equation}
with the orbital velocity $v_{\rm orb} = \sqrt{G M_\star / R_\star}$. Although our catalog is based on Table A1 of \cite{shultz2022}, there may be systematic deviations due to the derived quantities. In particular, the expression of the CBO luminosity includes the stellar radius $R_\star$, which we estimate from the Stefan-Boltzmann law given the bolometric luminosities and effective temperatures. The angular velocity is calculated from the rotation periods as $\Omega=2 \pi / P_{\rm rot}$. 
A linear regression fit to the data from previously detected sources yields a best-fit line with a slope of $m=0.88$ and an offset of $b=-8.52$. This fit is consistent with the theoretical prediction that \( L_{\rm radio} \) scales with \( L_{\rm CBO} \), modulated by a \( 10^{-8} \) empirical factor to account for energy losses in the magnetosphere. The purple line represents this theoretical scaling, while the shaded grey regions depict the 1 and 3~\(\sigma\) uncertainties around the best-fit regression.

Comparing the spectral radio luminosity to the CBO luminosity yields the same slope but -- due to the different units -- a different offset (Figure \ref{fig:lcbolrad2}). Here, for simplicity, we calculated $L_{\rm radio, \nu}$ from the radio luminosities reported in Table A1 of \cite{shultz2022} and scaled them by $\Delta \nu$ (Equation~\ref{eq:2}). This diagram is also consistent with the works of \cite{leto2021,leto2022}.

The coefficient of determination, \( R^2 = 0.68 \), indicates that 68\% of the variance in \( \log(L_{\rm radio}) \) can be explained by the linear relationship with \( \log(L_{\rm CBO}) \). While this suggests a strong correlation, the remaining 32\% of variance may arise from intrinsic scatter due to differences in stellar properties (e.g., magnetic topology, wind self-absorption) or measurement-related uncertainties (e.g., flux-to-luminosity conversion assumptions).
A Kolmogorov-Smirnov (KS) test performed on the residuals yields a \( p \)-value of 0.07, suggesting that the residuals are marginally consistent with a normal distribution at the 5\% significance level. This result supports the validity of the linear model but highlights some deviations, which may point to systematic effects or additional physical processes not captured by the CBO model. These deviations are particularly evident in non-detections with low \( L_{\rm radio} \). Interestingly, a large scatter also concerns those previous detections that have high  \( L_{\rm CBO} \). We will return to address this later.

The four out of five stars we observed resided above the linear regression as previous non-detections. 
In the case of HD~22470, our current measurement has similar sensitivity as the previous observation, yielding a 3~$\sigma$ upper limit of $\log L_{\rm radio}/L_\odot \approx -7$ for the non-detection. Provided that the rotational and magnetic parameters are accurate, this confirms that HD~22470 is outside of the 1$\sigma$ correlation, although the uncertain shape of the SED also plays an important role in this interpretation.
The non-detections reported for CPD-271791, HD36668, and HD49333 imply lower radio luminosities than expected from the CBO model, using the $10^{-8}$ scaling factor. However, if the true scaling factor varies significantly between stars due to magnetospheric or wind properties, the inferred CBO luminosity may overestimate the expected radio luminosity for stars with dense or self-absorbing magnetospheres. Additionally, the assumption of a flat SED shape in the 1-30 GHz range may not hold for stars with sub-GHz spectral turnovers or weaker gyrosynchrotron emission. Future studies incorporating multi-frequency data and refined scaling relations could help resolve these discrepancies.
Considering the empirical best-fit relation, these non-detections place HD36668 and HD49333 below the linear regression established for previous detections. To establish the compatibility of CPD-271791 with previous detections, its measurement sensitivity on \( L_{\rm radio} \) needs to be further improved by an order of magnitude. 

For HD~55522, we find a radio luminosity which is perfectly in line with the linear relation for previous detections (but similar to the previous detections, is below the theoretical CBO model prediction, assuming a scaling factor of $10^{-8}$). We conclude that this detection conforms to the CBO model, and thus strengthens the argument for a rotation-powered mechanism behind the gyrosynchrotron radio emission for rapidly rotating magnetic hot stars. At the same time, the non-detections also underscore the need for further refinement in theoretical modeling to account for outliers and scatter, particularly at lower luminosities.

\subsubsection{Scatter in the CBO - radio luminosity relation}

The lack of long-term optical spectropolarimetric monitoring and phase-resolved and multiband radio observations introduces additional uncertainties in the interpretation of both detections and non-detections. For instance, optical spectropolarimetric data are essential for accurately constraining the magnetic field geometry (e.g., obliquity angle, deviations from dipolar geometry), which directly affects the calculation of CBO luminosities. Without these data, stars with complex or poorly constrained magnetic fields may have misestimated $L_{\rm CBO}$ values, potentially explaining some of the outliers in the $L_{\rm CBO}$ vs. $L_{\rm radio}$ correlation.

Similarly, the absence of phase-resolved radio data complicates the assessment of non-detections. Radio luminosity from gyrosynchrotron emission is expected to vary with the rotational phase, and coherent auroral emission can appear as brief, high-flux pulses near magnetic nulls. A non-detection during a single phase may simply reflect observing at a flux minimum, rather than the absence of radio emission. Thus, phase-resolved observations are critical for confirming non-detections and identifying stars with transient or phase-dependent emission. We estimate that this translates to a factor of two uncertainty in the current CBO-radio luminosity relation since the expected variation of gyrosynchrotron emission between magnetic maximum and magnetic null is also about a factor of two. 

Multiband radio observations are critical for the SED assumption. Without knowing the peak flux of a target, the flux to luminosity conversion is simplified. This can lead to underestimating the radio flux, especially for observations conducted at the those frequencies where the SED is expected to decline (between 0.6 and 1.5 GHz and beteween 30 and 100 GHz, according to the \cite{shultz2022} SED assumption).

As evident in Figure~\ref{fig:lcbolrad}, previous detections can show a large range of radio luminosities inferred from flux measurements for a given theoretically-predicted CBO luminosity. 
We take the example of two previously detected radio bright stars, HD64740 and HD171247, which both have predicted CBO luminosities of $\log (L_{\rm CBO}/L_\odot)~=~3.0$ but their inferred radio luminosities are two orders of magnitude different, $\log (L_{\rm radio}/L_\odot) =$~$-7.19$ and $-5.16$, respectively. What causes this large difference?

In the case of HD64740, the detection is just within the 3~$\sigma$ uncertainty of the linear relation. A higher intrinsic mass-loss rate could cause self-absorption within the magnetosphere. This is currently not included in the CBO model but is expected to play a role in the detectability of gyrosynchrotron radiation from O-type stars, where the winds are generally stronger than in B-type stars \citep{shultz2022}. 

To reconcile the location of HD171247 within the CBO-radio luminosity plane, a possibility may be that its magnetic field strength and rotation period are incorrectly estimated. Indeed, \cite{shultz2022} argue that the rotation period can be either 1~d or 4~d, leading to a large uncertainty. For this reason, we discuss the impact of rotation periods next.

%
%
%
%
\begin{figure}
    \centering
    \includegraphics[width=0.45\textwidth]{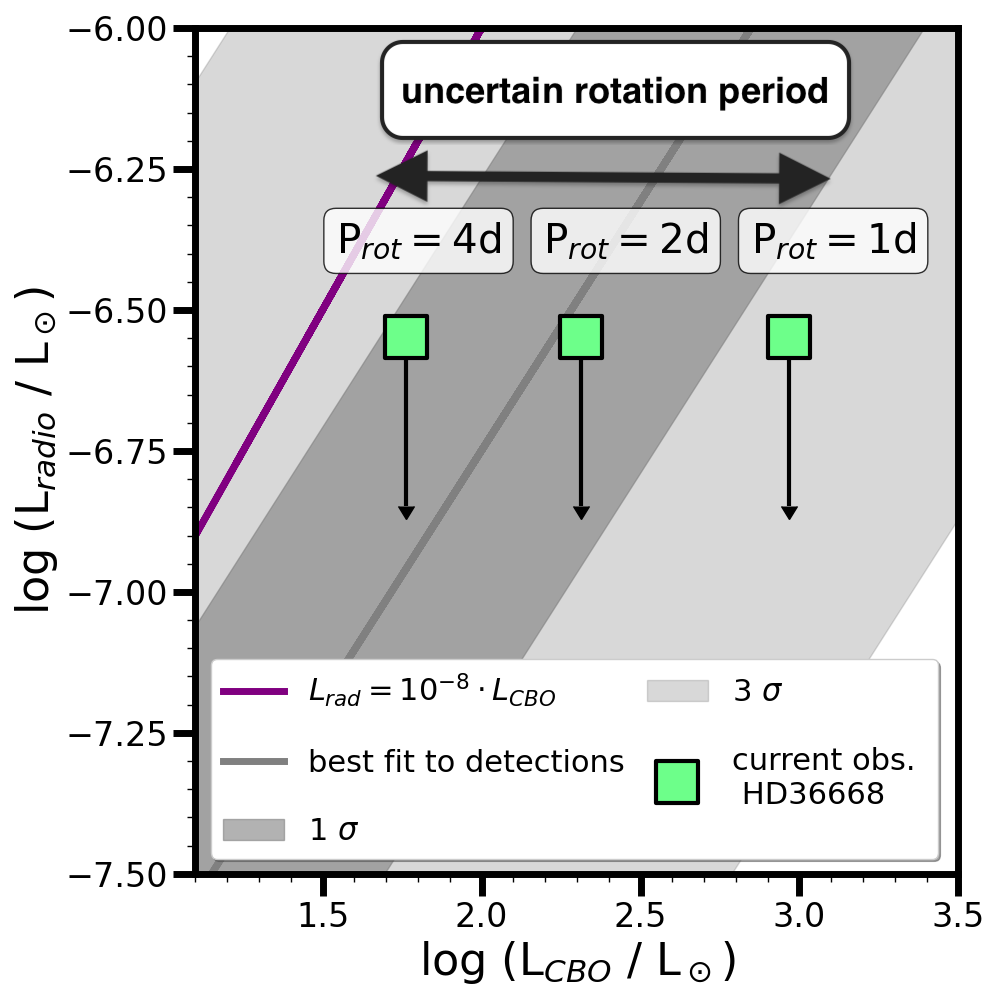}
    \caption{Radio luminosity versus the CBO luminosity. We use our flux density measurements to infer an upper limit on the radio luminosity of HD36668. A simple experiment demonstrates the impact of approx. a factor of two revision in the rotation period, shifting the CBO luminosity by a factor of four in one direction, and by a factor of 16 when considering both directions (towards lower and higher rotation periods than the adopted $P_{\rm rot}=2$~d). Such uncertainties could be one of the prime reasons for a large scatter in the CBO-radio luminosity relation.}
    \label{fig:prot2}
\end{figure}
%
%
%

%
%
\subsubsection{Impact of uncertain rotation periods}

Generally, the optical spectropolarimetric and the photometric monitoring provide robust constraints on the stellar rotation period. However, in some cases, due to incomplete or low-quality data, a coherent phasing cannot be firmly established. Stars with offset dipoles or more complex than dipole magnetic fields could also introduce deviations in the rotational phasing. 
In Figure~\ref{fig:prot2}, we demonstrate the consequences of relying on rotation periods that are not well-constrained. For example, for HD36668, \cite{shultz2022} primarily focused on the TESS photometry to constrain $P_{\rm rot} = 2.1192$~d. However, this rotation period still may not provide a satisfactory fit to the magnetic data.
Since the target stars are rapidly rotating, we take the assumption that a typical uncertainty in their rotation periods --if deviates largely from a coherent phasing-- could be roughly a factor of two. 
A factor of two uncertainty in the rotation period translates to a factor of four in the CBO luminosity. Thus --for the sake of this thought experiment-- if the rotation period of HD36668 was $P_{\rm rot} = 1$~d or 4~d, our measurement would imply that this star is outside of the 1$\sigma$ correlation or still above the linear regression, respectively.
Therefore, if the rotation periods are not well-constrained (but presumably within a factor of two for our primary selection of rapidly rotating magnetic stars with centrifugal magnetospheres), the predictive power of the CBO luminosity is only roughly an order of magnitude. This limitation can only be improved upon by monitoring campaigns, which provide accurate constraints on the rotational modulation.

%
%
%
%
\begin{figure*}
    \centering
    \includegraphics[width=0.45\textwidth]{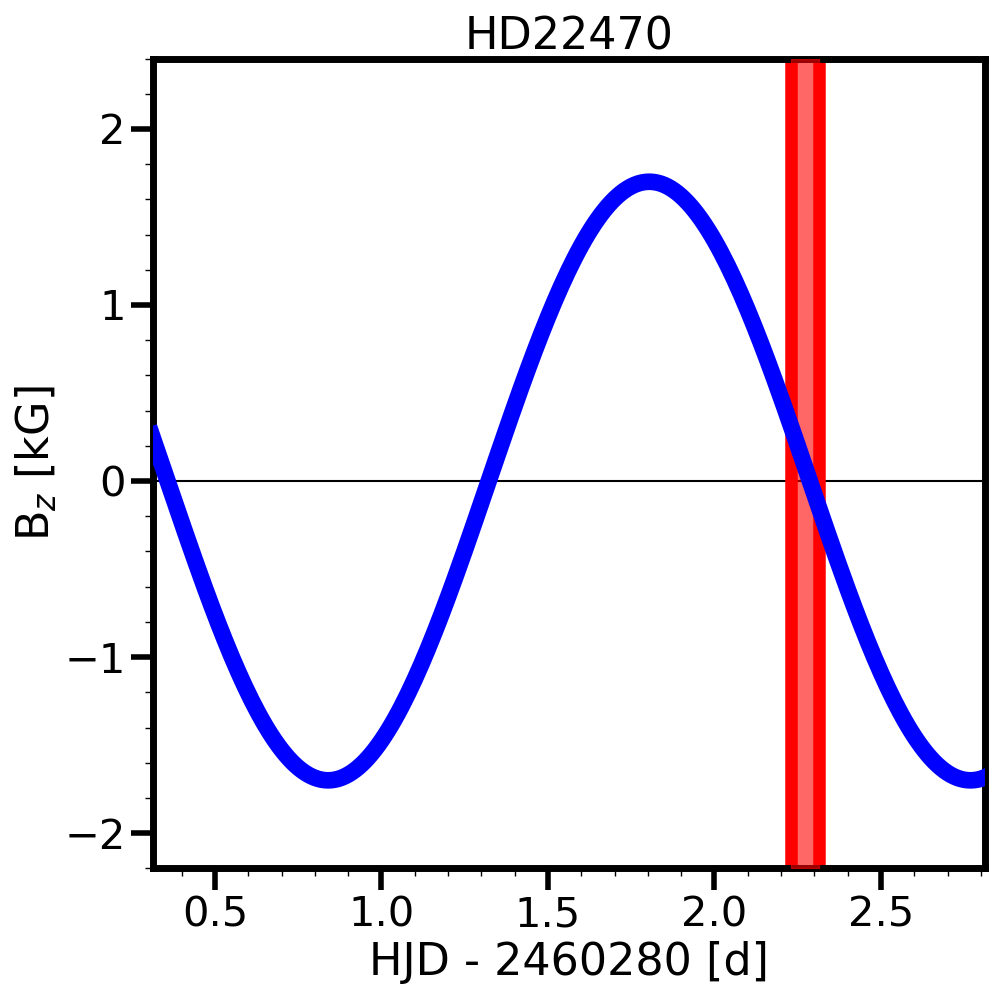}\includegraphics[width=0.45\textwidth]{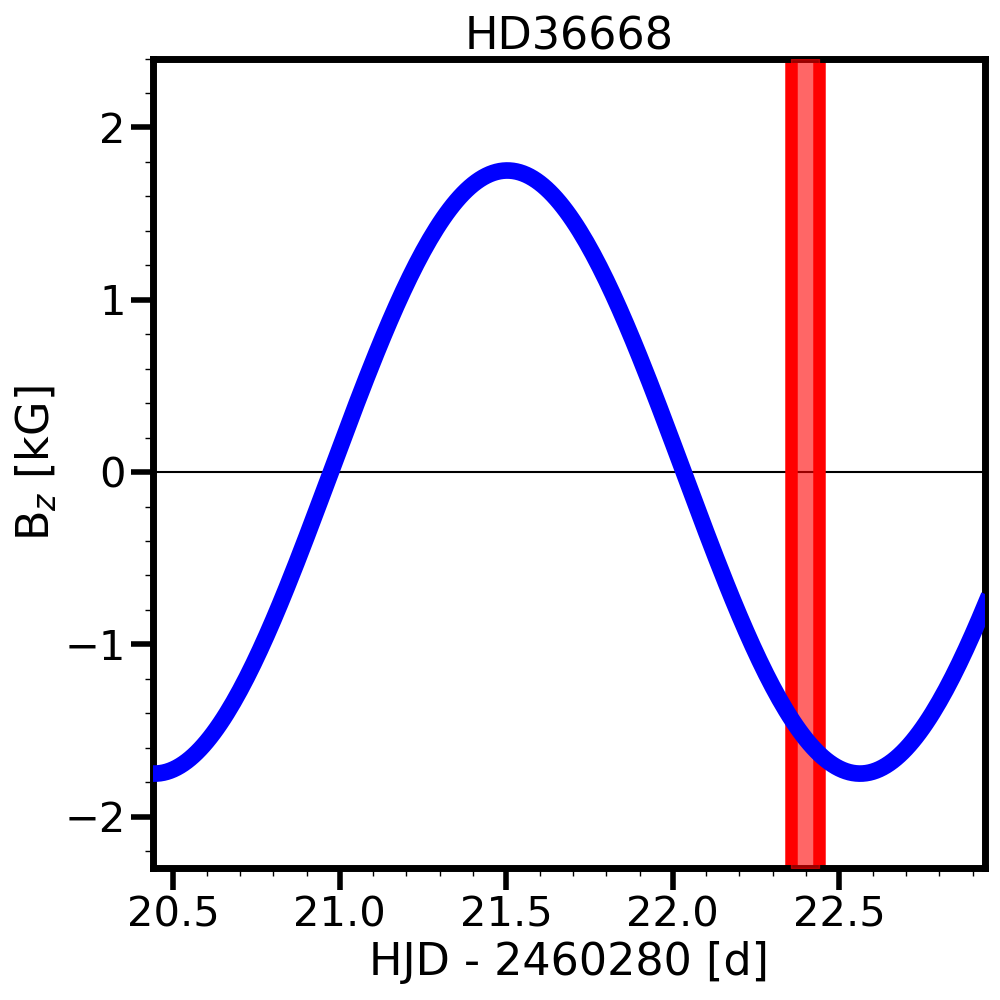}
    \includegraphics[width=0.45\textwidth]{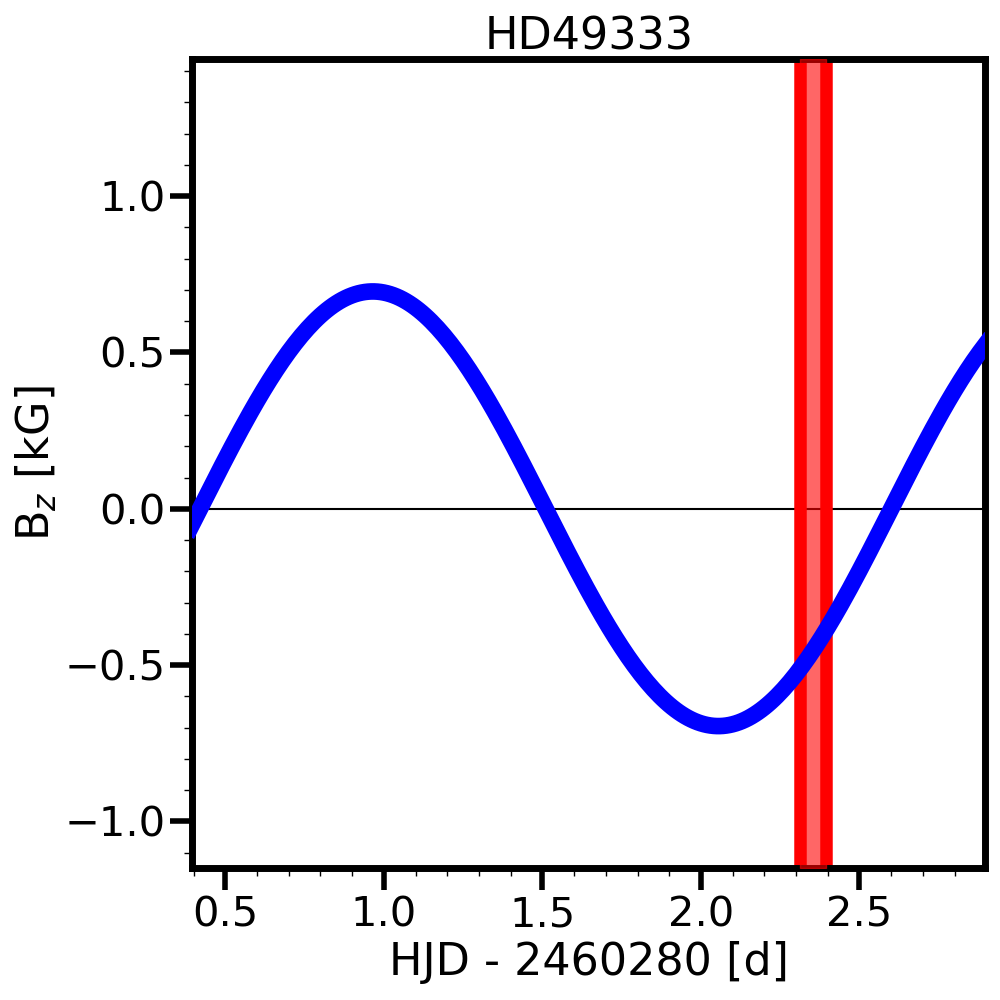}\includegraphics[width=0.45\textwidth]{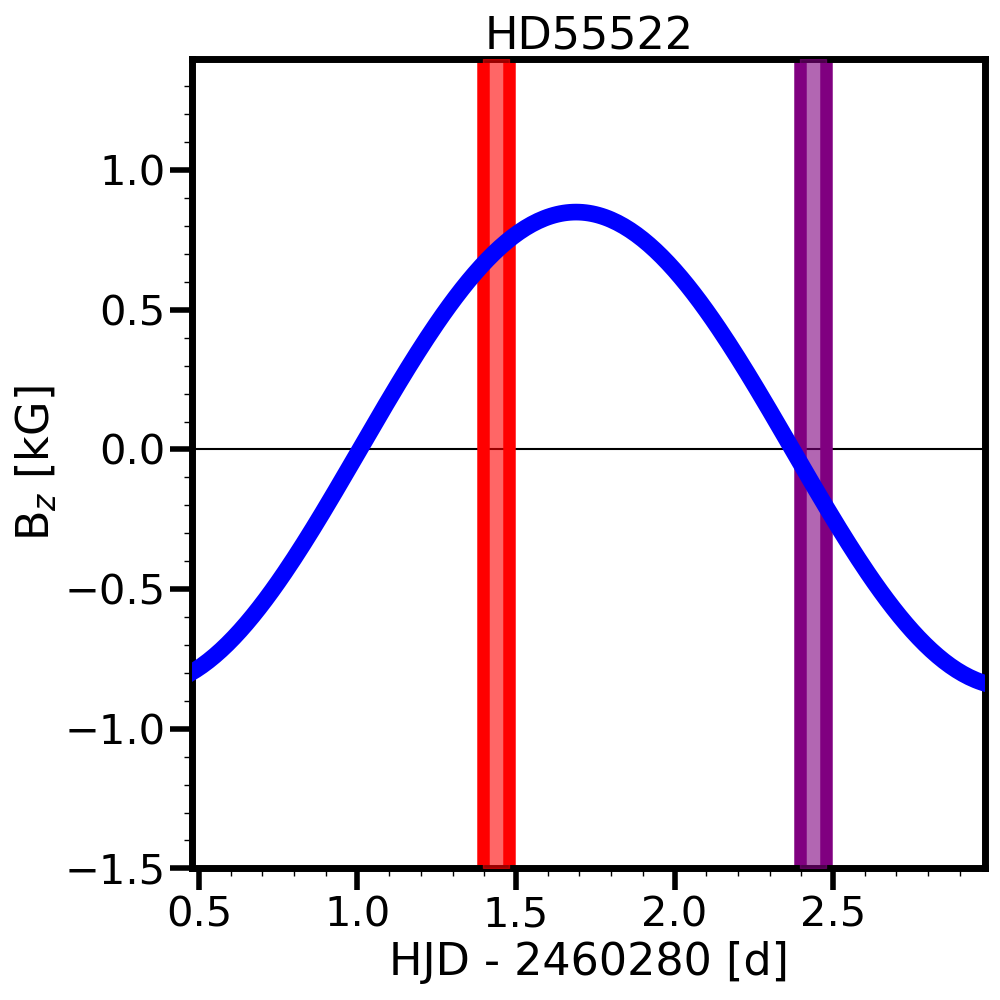}   
    \caption{The line-of-sight magnetic field strength as a function of the Heliocentric Julian Date is shown with a sinusoidal curve, which approximates the rotational modulation of a dipolar magnetic field. The timespan of our uGMRT radio observations is shown with the red (and purple) bands.}
    \label{fig:phases}
\end{figure*}
%
%
%
%
\subsection{Observed phases}

Figure \ref{fig:phases} illustrates the observed phases of four of the five stars in our current observational sample, showing the rotational modulation of their line-of-sight magnetic field ($B_z$). Observed phase curves are an essential diagnostic tool in magnetospheric studies of hot stars, as they provide insights into the geometry and temporal behavior of a star’s magnetic field over its rotation period, which are crucial for evaluating phenomena such as CBO and the origin of gyrosynchrotron emission.

For the present study, we were able to reconstruct phase curves for HD22470, HD36668, HD49333, and HD55522, based on the long-term spectropolarimetric monitoring data from \cite{bohlender1993}, \cite{shultz2020}, and \cite{shultz2022}. These observations allow us to map the rotational modulation of $B_z$ for each of these stars, which is presented as a sinusoidal curve. In contrast, CPD-271791 has not yet been subject to long-term monitoring, therefore its phase curve remains unconstrained. Establishing phase curves for all targets would be valuable in future campaigns to enable a full analysis of each star's magnetospheric behavior.

In each panel of Figure~\ref{fig:phases}, the sinusoidal curve represents the theoretical variation in $B_z$ expected from a dipolar magnetic geometry, modulated by the star’s rotation period. These are well matched by the previous spectropolarimetric data. The red and purple shaded regions indicate the time span of our uGMRT radio observations, providing constraints on the magnetic field alignment during those times. By comparing the radio flux observed at different phases with the corresponding $B_z$ values, we can begin to assess whether phase-dependent modulation of the magnetic field influences the radio emission characteristics. 

Our detection of radio emission from HD55522 is at a rotational phase close to magnetic maximum. On the other hand, we are unable to confirm detection at a subsequent rotation phase, at magnetic null. Since at the time of magnetic null the magnetic geometry has a preferential alignment to detect ECME, we inspected the data if a short burst of coherent emission could be identified. 

%
%
\subsection{Potential for ECMEs at magnetic nulls}

Recently, the flux peaks caused by auroral radio emission are actively being investigated \citep{das2022a,das2022b,das2022c,das2023,das2024, kajan2024}.
In most cases, ECME is expected at times of magnetic null, though in some cases the tendency is not so clear. Since two of our targets were observed at their magnetic nulls, we have further inspected the data, searching for signs of ECME. 

To obtain light curves for HD55522 (at both epochs) and for HD22470, the data was binned into shorter time intervals. We used WSclean to create images accordingly. Considering the number of visibility data points, the {\it -intervals-out} parameter was set to 10 for HD55522 (both at epoch 1 and 2) and to 8 for HD22470. It is important to note that the beam size varies due to differences in uv coverage for each dataset.

For each subset of the data, the peak flux within a circular region of 4.0 arcseconds in radius (approximately equivalent to the beam major axis, BMAJ) centered on the source was measured. Additionally, the standard deviation within a circular region of 30 arcseconds in a radius centered 2 arcminutes north of the source was calculated. The S/N was determined by taking the ratio of the peak flux to the standard deviation. For cases where the S/N exceeded 5, the measured peak flux was adopted. Otherwise, three times the standard deviation was used as an upper limit.  

Despite careful analysis, no clear ECME bursts were identified in the light curves of HD55522 or HD22470. The absence of detectable coherent emission could result from several factors. First, the emission might occur at frequencies outside our observational band. ECME often exhibits a narrow spectral range, and our observations, centered at 650 MHz, might have missed the emission if its peak lies at lower or higher frequencies. Second, the measurement sensitivity is lower than in case of the time-integrated data, potentially missing a fainter emission. 

This analysis underscores the importance of phase-resolved and multi-frequency observations to comprehensively characterize ECME in magnetic hot stars. Future studies should focus on continuous monitoring over entire rotation cycles and extend coverage to both lower and higher frequencies to account for variability in ECME. The absence of ECME in our observations does not rule out its occurrence in these stars, as magnetic nulls remain the most likely phases for such emission. 
Detecting ECME in more magnetic hot stars will provide critical insights into the conditions that drive this emission, such as the local electron density and magnetic reconnection processes in centrifugal magnetospheres. Expanding the sample size and observing strategies will be pivotal for establishing the prevalence and properties of ECME \citep{das2022b}. We obtained VLA and MeerKAT observations over the rotation cycle of two stars, which will be reported in forthcoming publications (Sakemi et al., in prep.).

\subsection{Measurement sensitivity}

The rms noise levels obtained from our data typically reach tens of $\mu$Jy~beam$^{-1}$, which is consistent with the noise level expected from the integration time. However, the image quality is not good due to sparse uv-coverage, which results in incomplete sampling of spatial frequencies and introduces artifacts in the synthesized beam. These distortions can appear as patterns around bright sources, often resembling concentric rings or streaks, depending on the coverage and weighting scheme used.  Despite the use of techniques such as source peeling and other direction-dependent calibrations to account for ionospheric distortions and antenna-based errors, we find that completely mitigating contamination from bright sources remains challenging, particularly for sources located near the target in the field of view. This contamination can elevate the local noise floor and obscure faint emissions, limiting our ability to detect weak radio signals from the targets.

In particular, some targets have nearby bright radio sources in the field of view, which limits their detectability (Figures \ref{fig:star1_loc_rad}-\ref{fig:star4_loc_rad}). CPD-271719 and HD36668 have nearby uncataloged radio bright sources. The angular separation is large, clearly implying that these sources cannot be associated with the star.
HD49333 has the nearby radio source NVSS J064706-210033 \citep{stein2021}, and several uncataloged radio bright sources. One nearly overlapping source in particular could strongly affect the detectability.
In the case of HD22470, the data quality was severely impacted by RFI, leading to a narrow-band measurement, in contrast to the more favorable wide-band measurements. In this case, the previous upper limit (obtained from higher frequency observations) could not be improved. It is worth revisiting this target with improved sensitivity and data quality because the obtained upper limits imply a notably lower radio luminosity than expected from the CBO relation. 

For HD55522, we cannot confirm radio detection at the second epoch of observations. The data quality is broadly similar to the one obtained at the first epoch, and the same integration time was used. This suggests that the non-detection at the second epoch is perhaps more likely caused by variability in the emission. Indeed, we deem that this detection is just missed by the lower gyrosynchrotron emission due to the observed phase of the star at the second epoch (magnetic null).

These limitations suggest that our non-detections are most likely due to insufficient measurement sensitivity rather than the absence of radio emission. Achieving higher signal-to-noise ratios for these targets, particularly in fields with strong source contamination, will require longer integration times and improved observational setups, such as MeerKAT or the Square Kilometre Array (SKA).

%
%
\subsection{Future prospects with SKA}

The SKA offers an unparalleled opportunity to advance our understanding of the magnetospheres of massive stars. With its unprecedented sensitivity, broad frequency coverage, and rapid survey speed, the SKA will significantly enhance the detection and characterization of non-thermal radio emission from magnetic OB stars \citep{dewdney2009,braun2015}. 

The SKA’s wide frequency range, spanning from 50 MHz to 15 GHz across its Low and Mid components, is critical for constraining the SED of magnetic hot stars. Cross-calibrated, multi-frequency observations from the SKA will refine models of particle acceleration and allow for testing predictions such as the spectral turnover and the flat SEDs assumed in scaling relations \citep{shultz2022}. Such observations will also provide insights into the physical conditions in the magnetosphere, such as plasma density, magnetic field topology, and loss mechanisms.

For HD55522, which displays a gyrosynchrotron flux density of ~160 $\mu$Jy at 650 MHz, the SKA1-Mid would achieve a signal-to-noise ratio of ~70 in just 10 minutes, compared to 40 minutes with the uGMRT. 
The SKA's sensitivity will allow the detection of similar emissions from stars up to 10 times farther away, effectively expanding the observable sample volume by a factor of 1000. Additionally, for stars with lower radio luminosities, such as those with non-detections in the current study, the SKA would lower flux limits by more than an order of magnitude. This improved sensitivity will also allow for detecting emission from stars with dense, self-absorbing magnetospheres\footnote{\url{https://www.skao.int/en/science-users/118/ska-telescope-specifications}}. 
The SKA's wide field of view (FOV), about 2.5 times larger than that of uGMRT at the 650 MHz band, will also be an advantage in increasing the number of observations of stellar sources, by the commensal use of other programs. Although we can only measure the flux density from a snapshot image, a source may be serendipitously observed multiple times, if it is within the FOV of other programs. Utilizing these observations, an improved rotational phase coverage can be accomplished for radio stars.
These capabilities will not only solidify scaling relations like the CBO-radio luminosity correlation but also extend them to a broader parameter space, encompassing more diverse stellar populations \citep{owocki2022, shultz2022}.

The SKA’s ability to push the boundaries of sensitivity, frequency coverage, and larger FOV will allow for a much improved statistical exploration of magnetic massive stars, revealing how their radio emission evolves with stellar parameters such as rotation, magnetic field strength, and mass-loss rates. By enabling detailed studies of individual systems and population-level trends, the SKA will bridge gaps between theory and observation, significantly advancing our understanding of magnetospheric processes and their implications for massive star evolution.

\subsection{Broader implications for magnetospheric physics}

Expanding radio observations to a larger sample of magnetic massive stars could refine our understanding of how magnetospheric properties vary across different stellar parameters, such as magnetic field strength, rotation rate, and stellar wind density. 

Our findings at radio frequencies underscore the need for a multiwavelength approach to fully understand magnetospheric processes in massive stars. In particular, X-ray emission has been reported in previous studies \citep[e.g.,][]{babel1997,naze2014}. 
Combining radio data with X-ray observations could reveal complementary insights into particle acceleration mechanisms, plasma density, and magnetic reconnection events within stellar magnetospheres. High-energy X-ray follow-up observations can also test for hot plasma signatures associated with shocks or magnetic reconnection, which are anticipated by the CBO model.

\section{Conclusions}
\label{sec:concl}

The RAMBO project aims to explore the non-thermal radio emission from massive magnetic stars, particularly those with centrifugal magnetospheres where rotation-driven processes are predicted to generate gyrosynchrotron radiation. In the first campaign of the project, we employed the upgraded Giant Metrewave Radio Telescope to search for radio signatures in a selected sample of stars, whose rotation periods are less than five days. Our observations have yielded the first confirmed detection of radio emission from HD55522 at 650 MHz, supporting its classification as a new radio-bright magnetic hot star. This detection from HD55522 aligns with predictions from the CBO model, supporting the role of stellar rotation as a mechanism powering the gyrosynchrotron radio emission. 
The interpretation of non-detections underscores the importance of addressing uncertainties in the CBO scaling factor, SED assumptions, and magnetic field geometries. Future efforts should prioritize multi-frequency campaigns and refine theoretical modeling to reduce these uncertainties and improve the predictive power of the CBO model.

Future steps for the RAMBO project include expanding our sample across a broader frequency range to capture a more comprehensive spectrum of radio emission and conducting rotational phase-resolved observations to assess variability. In addition, undertaking complementary X-ray and optical studies to investigate multi-wavelength signatures of magnetospheric activity would be logical extensions. Ultimately, these efforts will provide deeper insight into the origin of radio emission from ordered magnetospheres.

\section*{Data Availability Statement}

The data is publicly available from the 
GMRT Online Archive, \url{https://naps.ncra.tifr.res.in/goa/data/search}. Intermediate and calibrated data products are available from the authors upon reasonable request.

\section*{Acknowledgements}

We sincerely thank the anonymous referee for their constructive comments, which contributed to improving the manuscript.
Z.K. acknowledges support from JSPS Kakenhi Grant-in-Aid for Scientific Research (23K19071).
H.S. acknowledges support from JSPS Kakenhi Grant-in-Aid for Scientific Research (22K20386).
We thank the staff of the GMRT that made these observations possible. GMRT is run by the National Centre for Radio Astrophysics of the Tata Institute of Fundamental Research.

%
%
\bibliography{main}{}
\bibliographystyle{aasjournal}


\section*{Appendix}

\begin{figure*}[h!]
    \centering
    \includegraphics[width=0.45\textwidth]{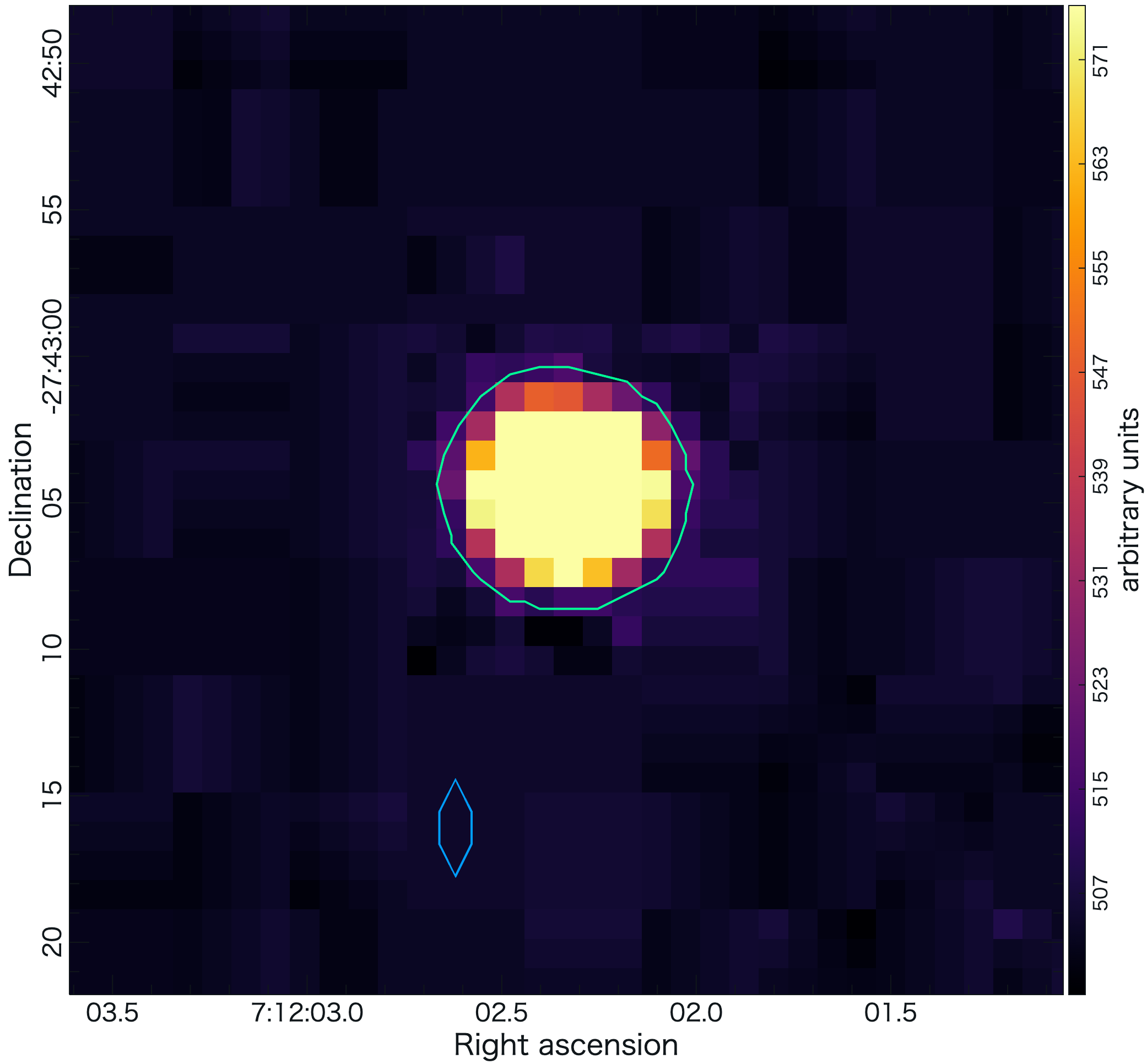}\includegraphics[width=0.45\textwidth]{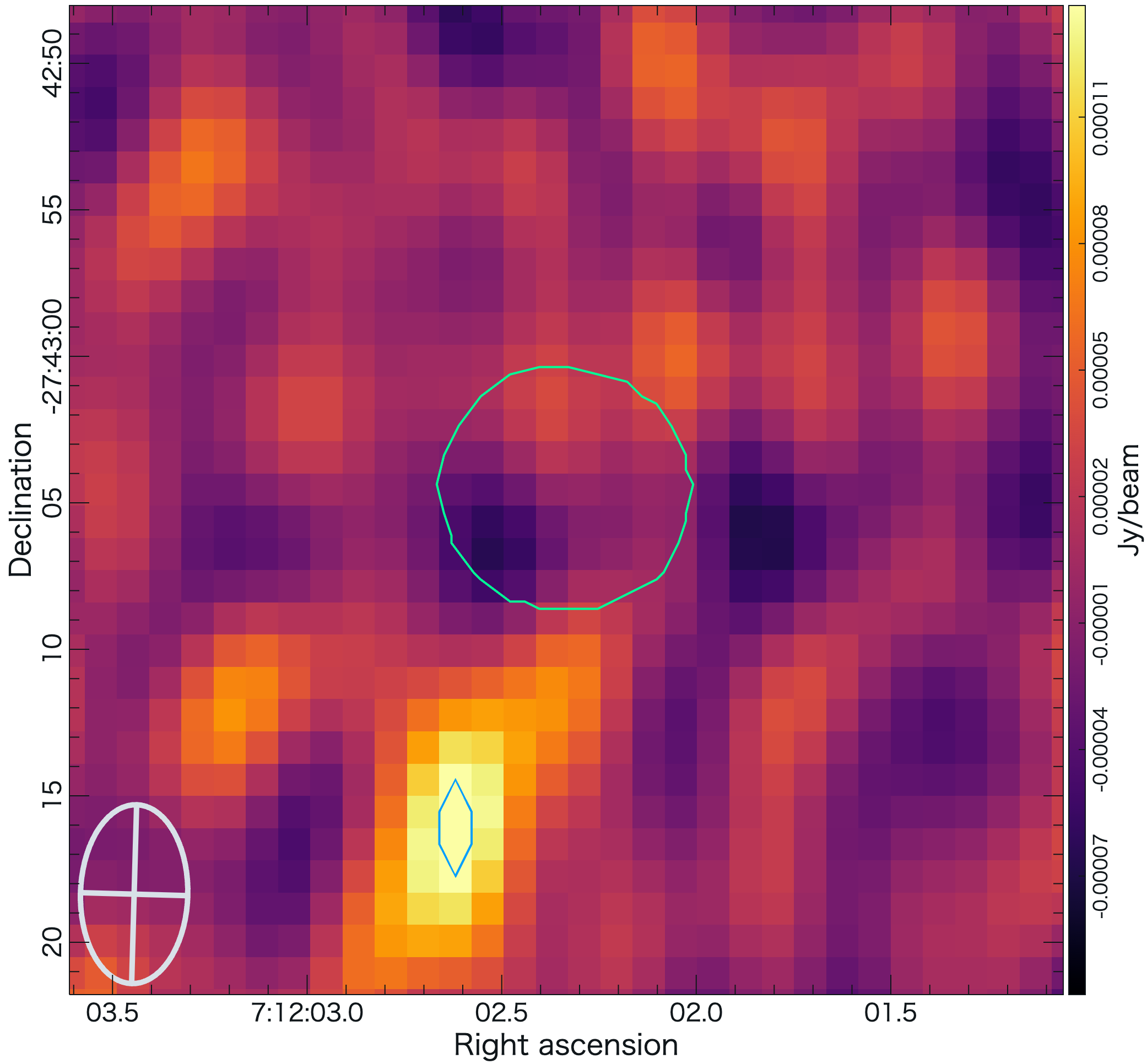}   
    \caption{CPD-271791. Left: K-band infrared image from 2MASS. Right: our uGMRT observations.}
    \label{fig:star1_loc_rad}
\end{figure*}

\begin{figure*}[h!]
    \centering
    \includegraphics[width=0.45\textwidth]{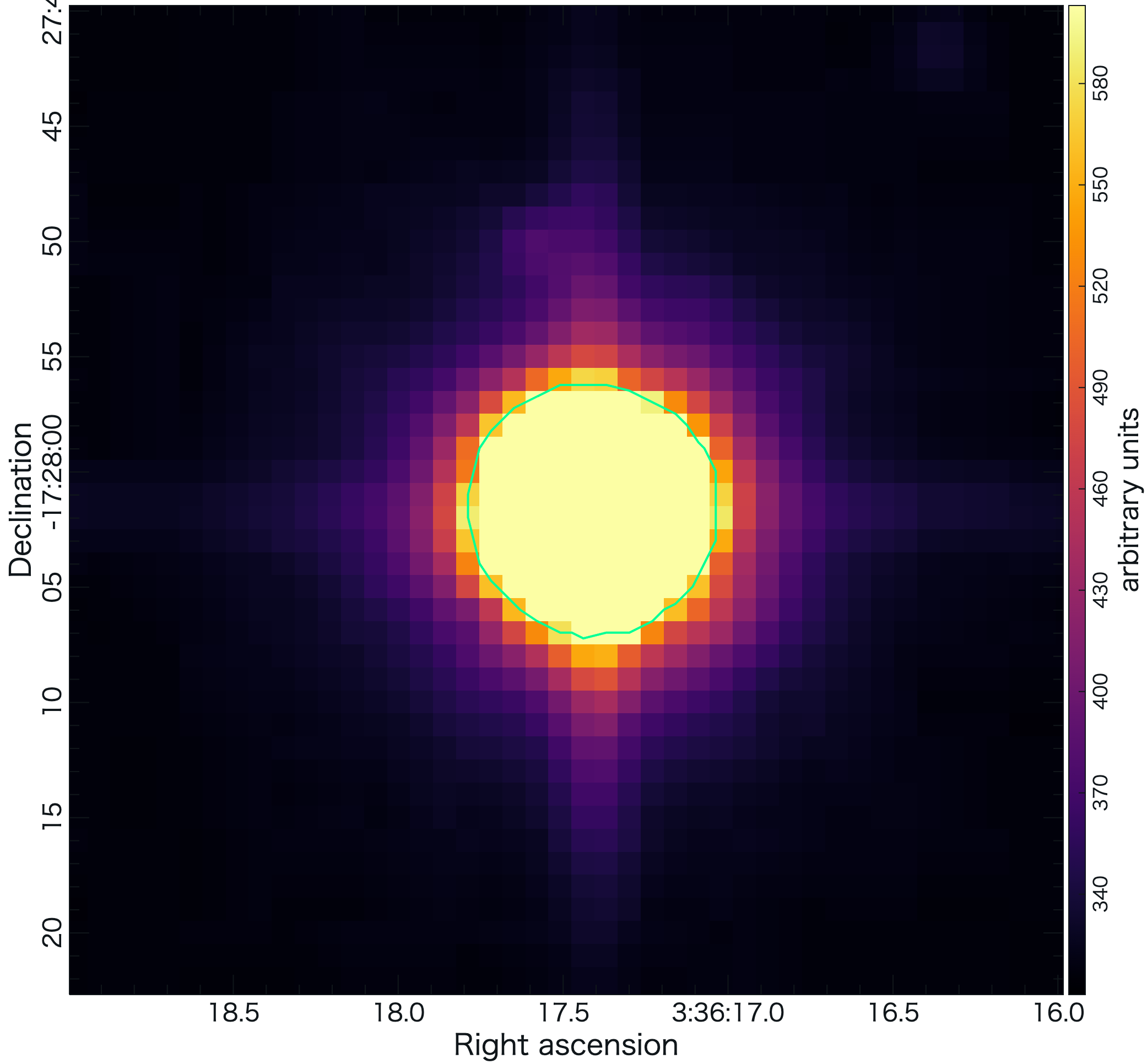}\includegraphics[width=0.45\textwidth]{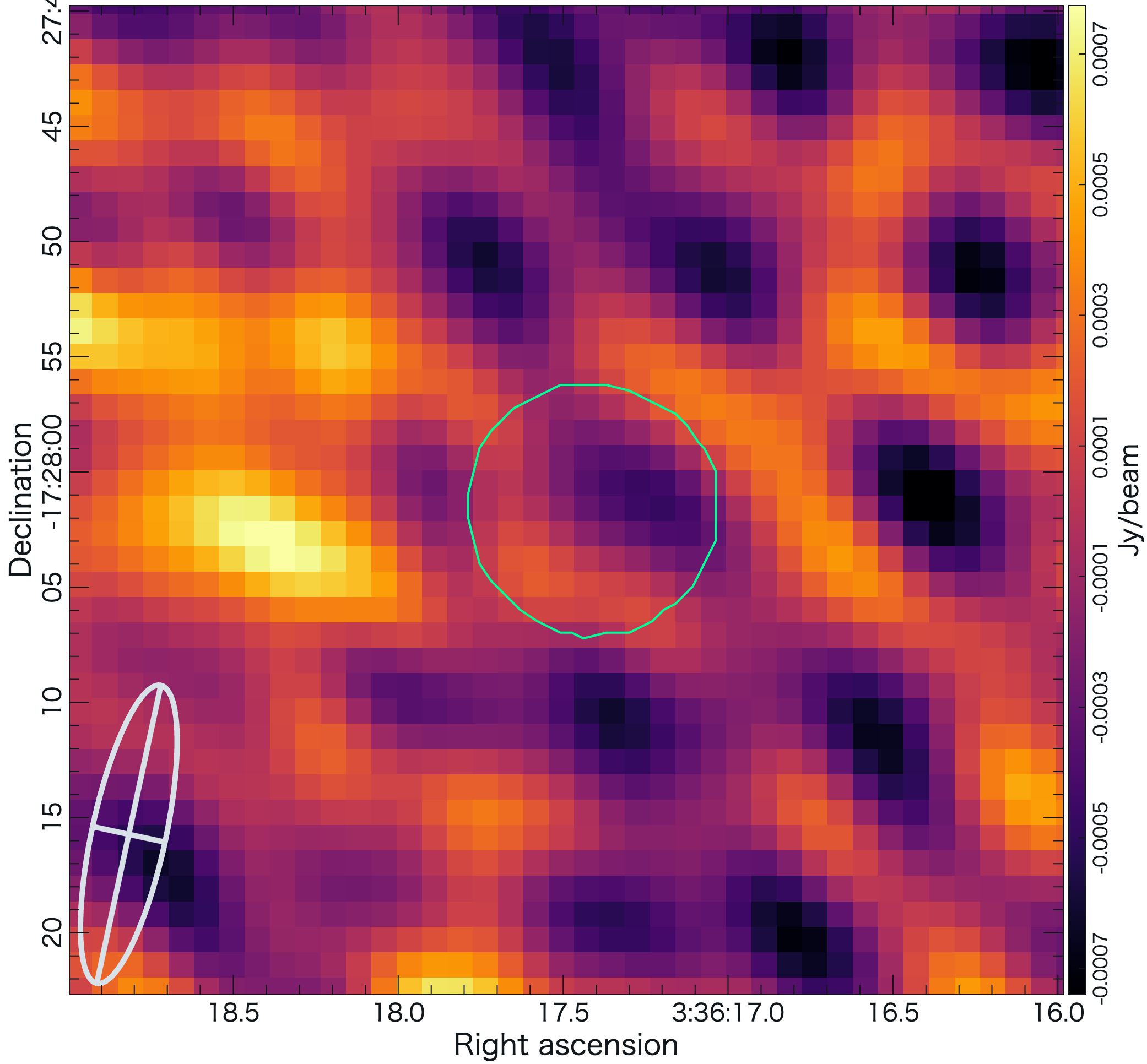}   
    \caption{HD22470. Left: H-band infrared image from 2MASS (since the K-band image is not available). Right: our uGMRT observations.}
    \label{fig:star2_loc_rad}
\end{figure*}

\begin{figure*}[h]
    \centering
    \includegraphics[width=0.45\textwidth]{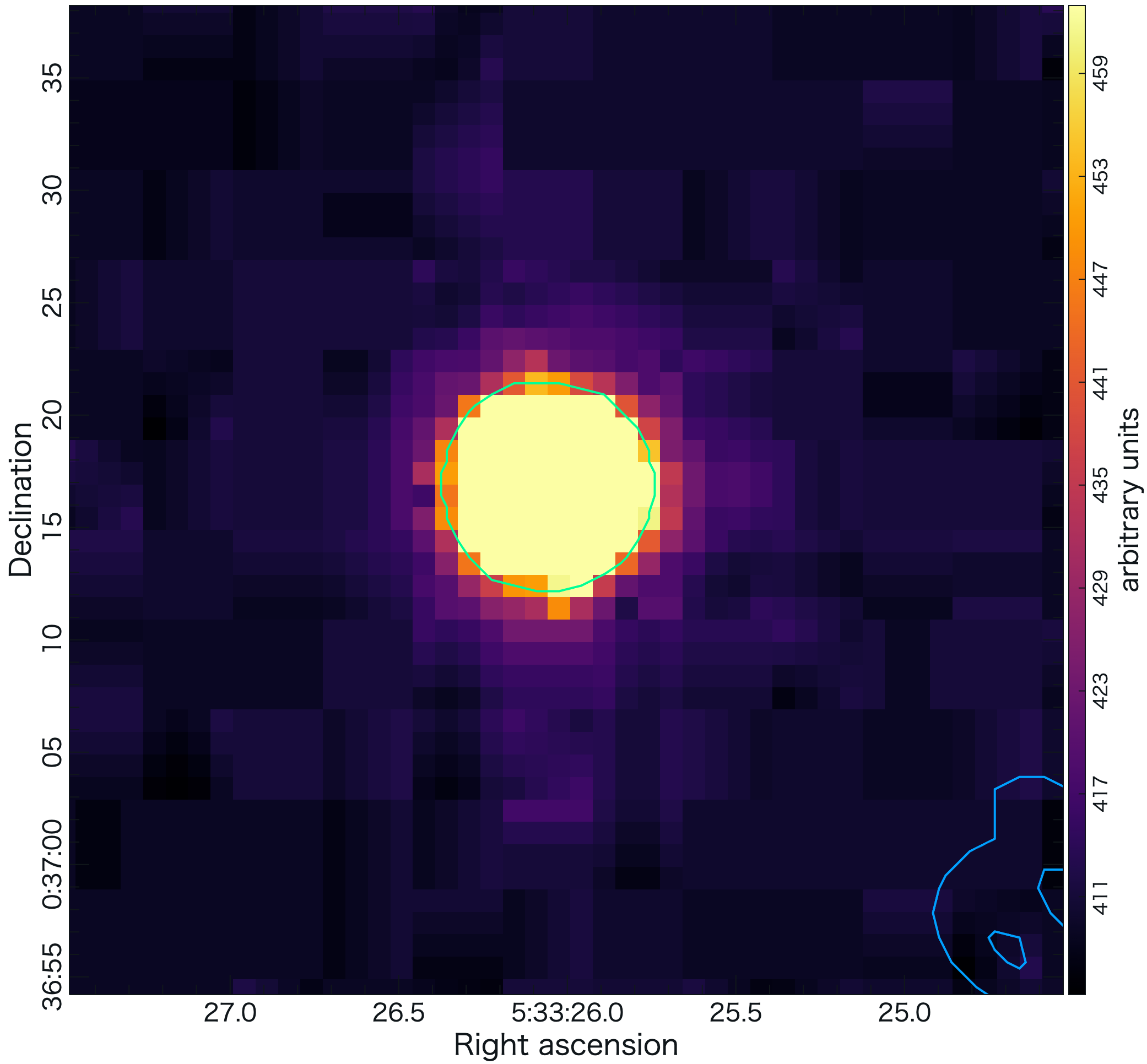}\includegraphics[width=0.45\textwidth]{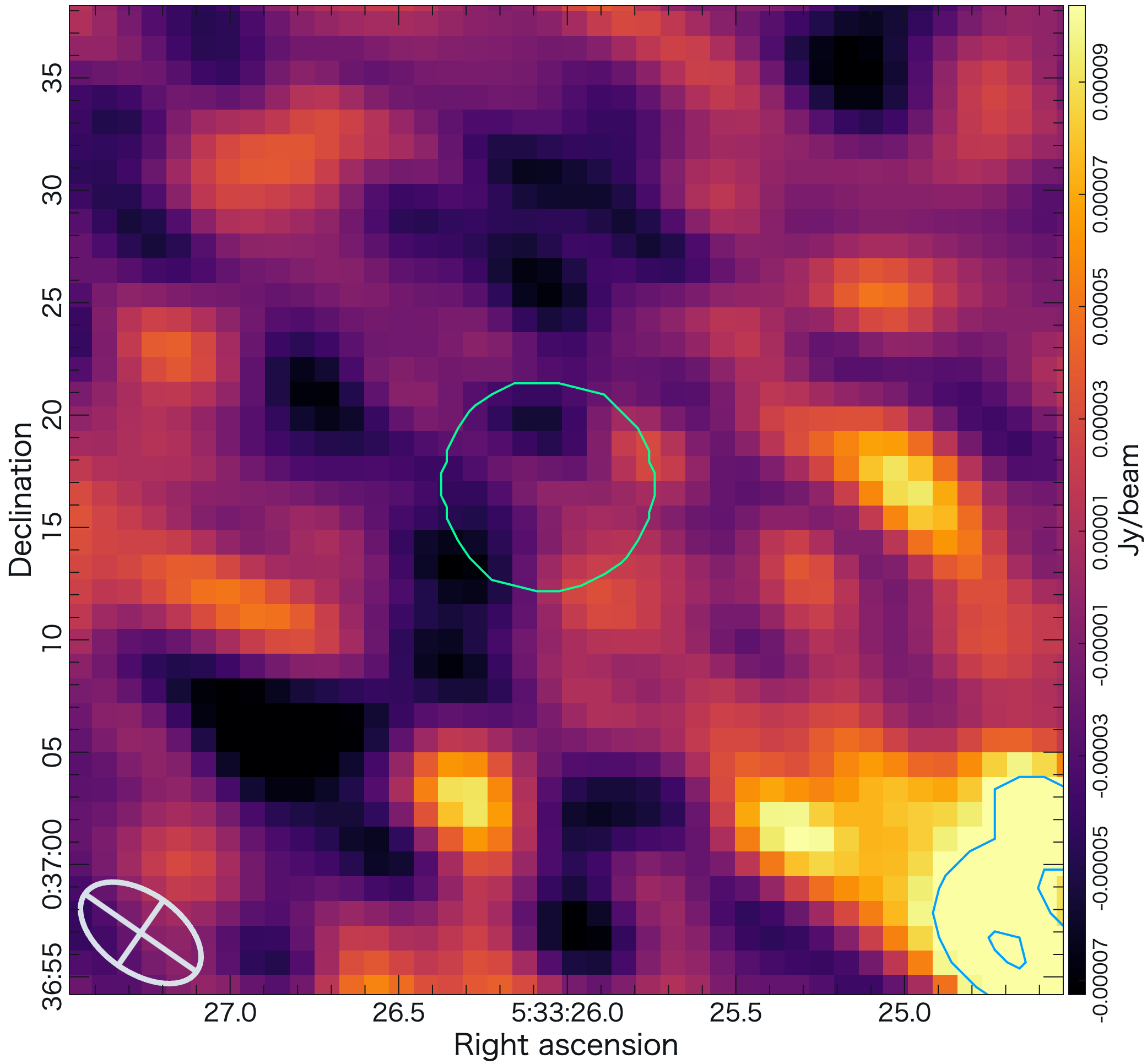}   
    \caption{HD36668. Left: K-band infrared image from 2MASS. Right: our uGMRT observations.}
    \label{fig:star3_loc_rad}
\end{figure*}

\begin{figure*}[h]
    \centering
    \includegraphics[width=0.45\textwidth]{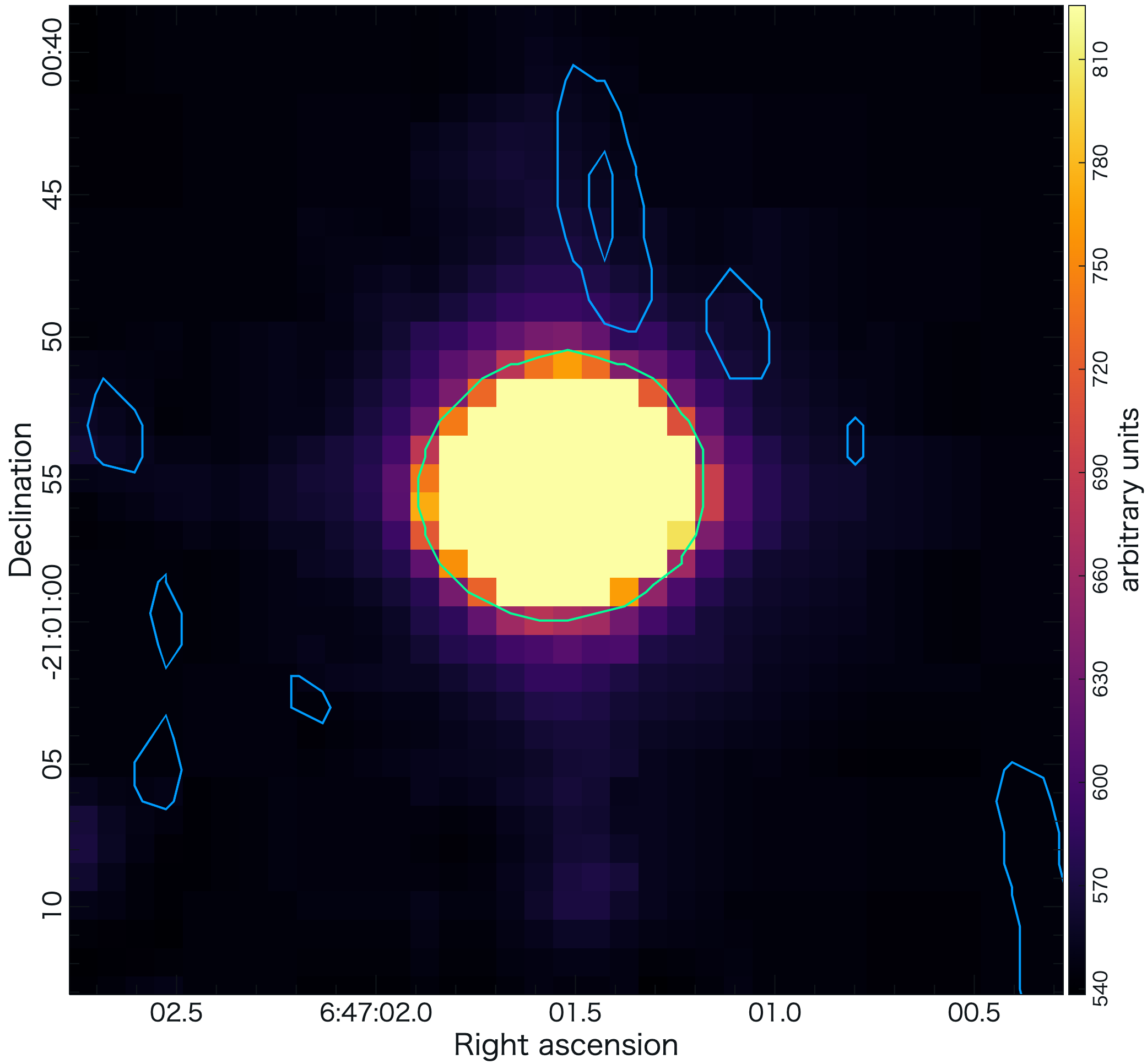}\includegraphics[width=0.45\textwidth]{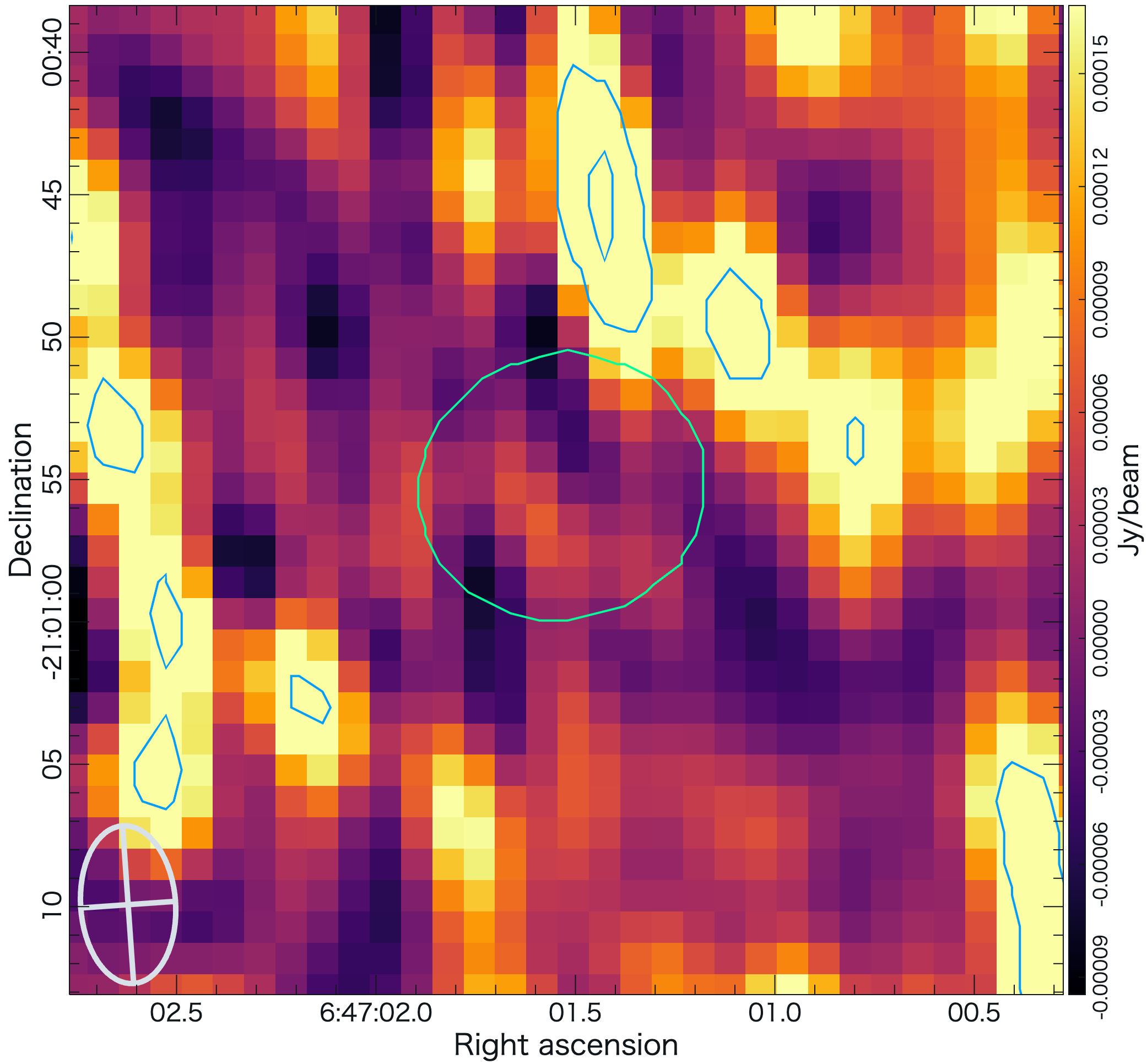}   
    \caption{HD49333. Left: K-band infrared image from 2MASS. Right: our uGMRT observations.}
    \label{fig:star4_loc_rad}
\end{figure*}

\begin{figure*}[h]
    \centering
    \includegraphics[width=0.45\textwidth]{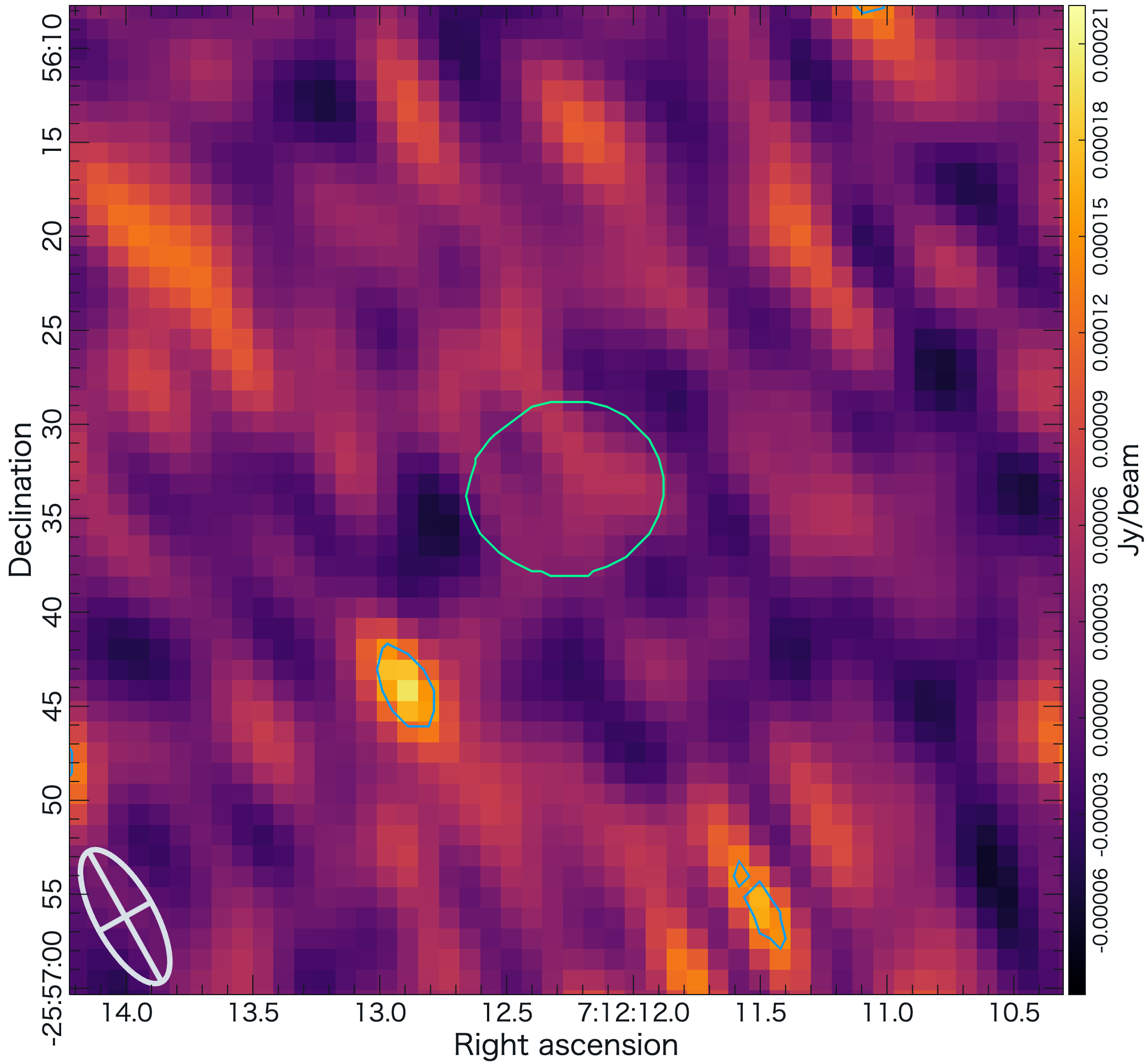}  
    \caption{HD55522. Our uGMRT observation from Dec 5, 2023.}
    \label{fig:star5_loc_rad}
\end{figure*}

Figures \ref{fig:star1_loc_rad} - \ref{fig:star5_loc_rad} show the radio images of the non-detections.

%
%

%
%
%

\end{document}